\documentclass[10pt,sigconf,nonacm]{acmart}

\usepackage{amsmath}
\usepackage[capitalize]{cleveref}
\usepackage{subcaption}
\usepackage{booktabs}
\usepackage{tabularx}
\usepackage{tikz}
\usetikzlibrary{fpu,calc,positioning,trees,shapes,matrix,intersections,fit}
\usepackage{pgfplots}
\usepackage{pgfplotstable}
\pgfplotsset{compat=1.17}
\usepgfplotslibrary{groupplots}
\usepackage[binary-units]{siunitx}
\usepackage{comment}
\usepackage{mathtools}

\usepackage{mathdots}

\pgfplotstableread{timedata-2/lammps-prof.dat}\lammpsProf
\pgfplotstablemodifyeachcolumnelement{threads}\of\lammpsProf\as\cell{%
  \ifthenelse{\equal{\cell}{nan}}{
    \edef\cell{--}
  }{
    \pgfmathsetmacro{\nodes}{int(round(\cell / 42))}
    \edef\cell{\cell{}t}
  }
}
\pgfplotsforeachungrouped \col in {prof1_time,prof2_time} {
  \pgfplotstablemodifyeachcolumnelement{\col}\of\lammpsProf\as\cell{%
    \ifthenelse{\equal{\cell}{nan}}{
      \edef\cell{--}
    }{
      \ifthenelse{\equal{\cell}{timeout}}{
        \edef\cell{$>$1hr}
      }{
        \ifthenelse{\equal{\cell}{noattempt}}{
          \edef\cell{NA}
        }{
          \ifthenelse{\equal{\cell}{oom}}{
            \edef\cell{O/M}
          }{
            \pgfmathsetmacro{\mins}{int(\cell / 60)}
            \pgfmathsetmacro{\secs}{int(round(mod(\cell , 60)))}
            \pgfmathsetmacro{\pad}{\secs < 10 ? "0" : ""}
            \edef\cell{\mins:\pad\secs}
          }
        }
      }
    }
  }
}
\pgfplotstabletranspose[columns={profiles,threads,prof1_time,prof2_time}]\lammpsProfT{\lammpsProf}

\pgfplotstableread{timedata-2/pelec-prof.dat}\pelecProf
\pgfplotstablemodifyeachcolumnelement{threads}\of\pelecProf\as\cell{%
  \ifthenelse{\equal{\cell}{nan}}{
    \edef\cell{--}
  }{
    \pgfmathsetmacro{\nodes}{int(round(\cell / 63))}
    \edef\cell{\cell{}t}
  }
}
\pgfplotsforeachungrouped \col in {prof1_time,prof2_time} {
  \pgfplotstablemodifyeachcolumnelement{\col}\of\pelecProf\as\cell{%
    \ifthenelse{\equal{\cell}{nan}}{
      \edef\cell{--}
    }{
      \ifthenelse{\equal{\cell}{timeout}}{
        \edef\cell{$>$1hr}
      }{
        \ifthenelse{\equal{\cell}{noattempt}}{
          \edef\cell{NA}
        }{
          \ifthenelse{\equal{\cell}{oom}}{
            \edef\cell{O/M}
          }{
            \pgfmathsetmacro{\mins}{int(\cell / 60)}
            \ifthenelse{\mins > 99}{
              \pgfmathsetmacro{\hours}{int(round(\cell / 60 / 60 * 10)) / 10}
              \edef\cell{\hours{}hr}
            }{
              \pgfmathsetmacro{\secs}{int(round(mod(\cell , 60)))}
              \pgfmathsetmacro{\pad}{\secs < 10 ? "0" : ""}
              \edef\cell{\mins:\pad\secs}
            }
          }
        }
      }
    }
  }
}
\pgfplotstabletranspose[columns={profiles,threads,prof1_time,prof2_time}]\pelecProfT{\pelecProf}

\pgfplotstablevertcat{\lammpsPelecProf}{\lammpsProf}
\pgfplotstablevertcat{\lammpsPelecProf}{\pelecProf}
\pgfplotstabletranspose[columns={profiles,threads,prof1_time,prof2_time}]\lammpsPelecProfT{\lammpsPelecProf}

\pgfplotstableread{timedata-2/amg-prof.dat}\amgProf
\pgfkeys{/pgf/fpu}
\pgfplotstablemodifyeachcolumnelement{threads}\of\amgProf\as\cell{%
  \ifthenelse{\equal{\cell}{nan}}{
    \edef\cell{--}
  }{
    \pgfmathsetmacro{\nodes}{int(round(\cell / 63))}
    \pgfmathprintnumberto[int detect,assume math mode]{\nodes}{\nodes}
    \edef\cell{\cell{}t}
  }
}
\pgfkeys{/pgf/fpu=false}
\pgfplotsforeachungrouped \col in {prof1_time,prof2_time,sc_time} {
  \pgfplotstablemodifyeachcolumnelement{\col}\of\amgProf\as\cell{%
    \ifthenelse{\equal{\cell}{nan}}{
      \edef\cell{--}
    }{
      \ifthenelse{\equal{\cell}{timeout}}{
        \edef\cell{$>$1hr}
      }{
        \ifthenelse{\equal{\cell}{noattempt}}{
          \edef\cell{NA}
        }{
          \ifthenelse{\equal{\cell}{oom}}{
            \edef\cell{O/M}
          }{
            \pgfmathsetmacro{\mins}{int(\cell / 60)}
            \pgfmathsetmacro{\secs}{int(round(mod(\cell , 60)))}
            \pgfmathsetmacro{\pad}{\secs < 10 ? "0" : ""}
            \edef\cell{\mins:\pad\secs}
          }
        }
      }
    }
  }
}
\pgfplotstabletranspose[columns={profiles,threads,sc_time,prof1_time,prof2_nglb_time,prof2_time}]\amgProfT{\amgProf}

\pgfplotstableread{timedata-2/lammps-app.dat}\lammpsApp
\pgfplotsforeachungrouped \col in {app_time} {
  \pgfplotstablemodifyeachcolumnelement{\col}\of\lammpsApp\as\cell{%
    \ifthenelse{\equal{\cell}{nan}}{
      \edef\cell{--}
    }{
      \ifthenelse{\equal{\cell}{timeout}}{
        \edef\cell{$>$1hr}
      }{
        \ifthenelse{\equal{\cell}{noattempt}}{
          \edef\cell{NA}
        }{
          \pgfmathsetmacro{\mins}{int(\cell / 60)}
          \pgfmathsetmacro{\secs}{int(round(mod(\cell , 60)))}
          \pgfmathsetmacro{\pad}{\secs < 10 ? "0" : ""}
          \edef\cell{\mins:\pad\secs}
        }
      }
    }
  }
}
\pgfkeys{/pgf/fpu}
\pgfplotsforeachungrouped \col in {ht1_space,prof1_space,ht2_space,prof2_space} {
  \pgfplotstablemodifyeachcolumnelement{\col}\of\lammpsApp\as\cell{%
    \ifthenelse{\equal{\cell}{nan}}{
      \edef\cell{}
    }{
      \pgfmathsetmacro{\gbs}{\cell / 1024}
      \pgfmathprintnumberto[1000 sep={},fixed relative,precision=3,assume math mode]{\gbs}{\gbs}
      \edef\cell{\gbs}
    }
  }
}
\pgfkeys{/pgf/fpu=false}

\pgfplotstableread{timedata-2/amg-app.dat}\amgApp
\pgfplotsforeachungrouped \col in {app_time} {
  \pgfplotstablemodifyeachcolumnelement{\col}\of\amgApp\as\cell{%
    \ifthenelse{\equal{\cell}{nan}}{
      \edef\cell{--}
    }{
      \pgfmathsetmacro{\mins}{int(\cell / 60)}
      \pgfmathsetmacro{\secs}{int(round(mod(\cell , 60)))}
      \pgfmathsetmacro{\pad}{\secs < 10 ? "0" : ""}
      \edef\cell{\mins:\pad\secs}
    }
  }
}
\pgfkeys{/pgf/fpu}
\pgfplotsforeachungrouped \col in {ht1_space,prof1_space,ht2_space,prof2_space,sp_space,sc_space} {
  \pgfplotstablemodifyeachcolumnelement{\col}\of\amgApp\as\cell{%
    \ifthenelse{\equal{\cell}{nan}}{
      \edef\cell{}
    }{
      \pgfmathsetmacro{\gbs}{\cell / 1024}
      \pgfmathprintnumberto[1000 sep={},fixed relative,precision=3,assume math mode]{\gbs}{\gbs}
      \edef\cell{\gbs}
    }
  }
}
\pgfkeys{/pgf/fpu=false}

\pgfplotstableread{timedata-2/pelec-app.dat}\pelecApp
\pgfplotsforeachungrouped \col in {app_time} {
  \pgfplotstablemodifyeachcolumnelement{\col}\of\pelecApp\as\cell{%
    \ifthenelse{\equal{\cell}{nan}}{
      \edef\cell{--}
    }{
      \ifthenelse{\equal{\cell}{timeout}}{
        \edef\cell{$>$1hr}
      }{
        \ifthenelse{\equal{\cell}{noattempt}}{
          \edef\cell{NA}
        }{
          \ifthenelse{\equal{\cell}{oom}}{
            \edef\cell{O/M}
          }{
            \pgfmathsetmacro{\mins}{int(\cell / 60)}
            \pgfmathsetmacro{\secs}{int(round(mod(\cell , 60)))}
            \pgfmathsetmacro{\pad}{\secs < 10 ? "0" : ""}
            \edef\cell{\mins:\pad\secs}
          }
        }
      }
    }
  }
}
\pgfkeys{/pgf/fpu}
\pgfplotsforeachungrouped \col in {ht1_space,prof1_space,ht2_space,prof2_space} {
  \pgfplotstablemodifyeachcolumnelement{\col}\of\pelecApp\as\cell{%
    \ifthenelse{\equal{\cell}{nan}}{
      \edef\cell{}
    }{
      \pgfmathsetmacro{\gbs}{\cell / 1024}
      \pgfmathprintnumberto[1000 sep={},fixed relative,precision=3,assume math mode]{\gbs}{\gbs}
      \edef\cell{\gbs}
    }
  }
}
\pgfkeys{/pgf/fpu=false}

\pgfplotstabletranspose[input colnames to={}, colnames from={threads},
  columns={threads,ht1_space,prof1_space,ht2_space,prof2_space,sp_space,sc_space,app_time}
]\amgProfSizes{\amgApp}
\pgfplotstableforeachcolumn{\amgProfSizes}\as\profs{%
  \pgfplotstablegetelem{0}{\profs}\of{\amgProfSizes} \edef\mhtOne{\pgfplotsretval}
  \pgfplotstablegetelem{1}{\profs}\of{\amgProfSizes} \edef\dhtOne{\pgfplotsretval}
  \pgfplotstablegetelem{2}{\profs}\of{\amgProfSizes} \edef\mhtTwo{\pgfplotsretval}
  \pgfplotstablegetelem{3}{\profs}\of{\amgProfSizes} \edef\dhtTwo{\pgfplotsretval}
  \pgfplotstablegetelem{4}{\profs}\of{\amgProfSizes} \edef\mScoreP{\pgfplotsretval}
  \pgfplotstablegetelem{5}{\profs}\of{\amgProfSizes} \edef\dScalasca{\pgfplotsretval}
  \edef\temp{\noexpand\pgfplotstableset{%
    create on use/amgM\profs/.style={create col/set list={--,In,\mScoreP,\mhtOne,--,\mhtTwo}},
    create on use/amgD\profs/.style={create col/set list={--,Out,\dScalasca,\dhtOne,--,\dhtTwo}},
  }}
  \temp
  \pgfplotstablegetelem{6}{\profs}\of{\amgProfSizes} \expandafter\edef\csname amgAppTime\profs\endcsname{\pgfplotsretval}
}

\pgfplotstabletranspose[input colnames to={}, colnames from={threads},
  columns={threads,ht1_space,prof1_space,ht2_space,prof2_space,app_time}
]\lammpsProfSizes{\lammpsApp}
\pgfplotstableforeachcolumn{\lammpsProfSizes}\as\profs{%
  \pgfplotstablegetelem{0}{\profs}\of{\lammpsProfSizes} \edef\mhtOne{\pgfplotsretval}
  \pgfplotstablegetelem{1}{\profs}\of{\lammpsProfSizes} \edef\dhtOne{\pgfplotsretval}
  \pgfplotstablegetelem{2}{\profs}\of{\lammpsProfSizes} \edef\mhtTwo{\pgfplotsretval}
  \pgfplotstablegetelem{3}{\profs}\of{\lammpsProfSizes} \edef\dhtTwo{\pgfplotsretval}
  \edef\temp{\noexpand\pgfplotstableset{%
    create on use/lammpsM\profs/.style={create col/set list={--,In,\mhtOne,\mhtTwo}},
    create on use/lammpsD\profs/.style={create col/set list={--,Out,\dhtOne,\dhtTwo}},
  }}
  \temp
  \pgfplotstablegetelem{4}{\profs}\of{\lammpsProfSizes} \expandafter\edef\csname lammpsAppTime\profs\endcsname{\pgfplotsretval}
}

\pgfplotstabletranspose[input colnames to={}, colnames from={threads},
  columns={threads,ht1_space,prof1_space,ht2_space,prof2_space,app_time}
]\pelecProfSizes{\pelecApp}
\pgfplotstableforeachcolumn{\pelecProfSizes}\as\profs{%
  \pgfplotstablegetelem{0}{\profs}\of{\pelecProfSizes} \edef\mhtOne{\pgfplotsretval}
  \pgfplotstablegetelem{1}{\profs}\of{\pelecProfSizes} \edef\dhtOne{\pgfplotsretval}
  \pgfplotstablegetelem{2}{\profs}\of{\pelecProfSizes} \edef\mhtTwo{\pgfplotsretval}
  \pgfplotstablegetelem{3}{\profs}\of{\pelecProfSizes} \edef\dhtTwo{\pgfplotsretval}
  \edef\temp{\noexpand\pgfplotstableset{%
    create on use/pelecM\profs/.style={create col/set list={--,In,\mhtOne,\mhtTwo}},
    create on use/pelecD\profs/.style={create col/set list={--,Out,\dhtOne,\dhtTwo}},
  }}
  \temp
  \pgfplotstablegetelem{4}{\profs}\of{\pelecProfSizes} \expandafter\edef\csname pelecAppTime\profs\endcsname{\pgfplotsretval}
}

\newcommand\selectProfilesNoclear[2]{%
  \pgfplotstableforeachcolumn{#1}\as\col{%
    \ifthenelse{\equal{\col}{colnames}}{}{%
      \pgfplotstablegetelem{0}{\col}\of{#1}
      \ifthenelse{\pgfplotsretval = #2}{
        \ifthenelse{\equal{\pgfkeysvalueof{/pgfplots/table/columns}}{}}{
          \edef\temp{\noexpand\pgfplotstableset{columns={\col}}}
          \temp
        }{
          \edef\temp{\noexpand\pgfplotstableset{columns/.add={}{,\col}}}
          \temp
        }
      }
    }
  }
}
\newcommand\selectProfiles[2]{%
  \pgfplotstableset{columns={}}%
  \selectProfilesNoclear{#1}{#2}%
}
\pgfplotstableset{
  create on use/lammpsLongSide/.style={create col/set list={--, --, {HPCToolkit~\cite{Toolkit:overview}}, {HPCToolkit-SA}}},
  create on use/amgLongSide/.style={create col/set list={--, --, {Scalasca~\cite{Scalasca}}, {HPCToolkit~\cite{Toolkit:overview}}, --, {HPCToolkit-SA}}},
  create on use/pelecLongSide/.style={create col/set list={--, --, {HPCToolkit~\cite{Toolkit:overview}}, {HPCToolkit-SA}}},
  create on use/lammpsSemilongSide/.style={create col/set list={--, --, {HPCToolkit~\cite{Toolkit:overview}}, {HPCToolkit-SA}}},
  create on use/amgSemilongSide/.style={create col/set list={--, --, {Scalasca~\cite{Scalasca}}, {HPCToolkit~\cite{Toolkit:overview}}, --, {HPCToolkit-SA}}},
  create on use/pelecSemilongSide/.style={create col/set list={--, --, {HPCToolkit~\cite{Toolkit:overview}}, {HPCToolkit-SA}}},
  create on use/nglbLammpsSide/.style={create col/set list={--, --, --, {w/o GLB}, {w/ GLB}}},
  create on use/nglbAmgSide/.style={create col/set list={--, --, --, --, {w/o GLB}, {w/ GLB}}},
}

\newcommand\amgNameEightK{AMG 8K}
\newcommand\amgNameSixteenK{AMG 16K}
\newcommand\amgNameSixFourK{AMG 65K}
\newcommand\amgNameTwoFiveSixK{AMG 262K}
\newcommand\lammpsNameOneK{LAMMPS 1K}

\newcommand\lammpsNameFourK{LAMMPS 4K}
\newcommand\pelecNameOneK{PeleC 1K}

\newcommand\pelecNameFourK{PeleC 4K}
\title{Preparing for Performance Analysis at Exascale}
\author{Jonathon Anderson}
\orcid{0000-0002-4506-2657}
\affiliation{%
  \institution{Rice University}
  \department{Department of Computer Science}
  \streetaddress{6100 Main St.}
  \city{Houston}
  \state{TX}
  \postcode{77005-1827}
  \country{USA}}
\email{janderson@rice.edu}

\author{Yumeng Liu}
\affiliation{%
  \institution{Rice University}
  \department{Department of Computer Science}
  \streetaddress{6100 Main St.}
  \city{Houston}
  \state{TX}
  \postcode{77005-1827}
  \country{USA}}
\email{Yumeng.Liu@rice.edu}

\author{John Mellor-Crummey}
\affiliation{%
  \institution{Rice University}
  \department{Department of Computer Science}
  \streetaddress{6100 Main St.}
  \city{Houston}
  \state{TX}
  \postcode{77005-1827}
  \country{USA}}
\email{johnmc@rice.edu}

\setcopyright{acmcopyright}
\copyrightyear{2022}
\acmYear{2022}
\acmConference[ICS'22]{International Conference on Supercomputing}{June 27--30}{Online}

\ccsdesc[500]{General and reference~Performance}
\ccsdesc[500]{Computing methodologies~Shared memory algorithms}
\ccsdesc[500]{Computing methodologies~Distributed algorithms}

\keywords{performance, parallelism}

\begin{abstract}
Performance tools for emerging heterogeneous exascale platforms must address two principal challenges when analyzing execution measurements.
First, measurement of large-scale executions may record mountains of performance data.
Second, performance measurements for parallel programs are sparse in two ways: the set of metrics present for any context and the set of contexts present in different threads.
For GPU-accelerated applications, an important source of sparsity is that none of the myriad of GPU metrics apply to any of the many CPU contexts. 
To address these challenges, we developed a novel \textit{streaming aggregation} approach to postmortem analysis that employs both shared and distributed memory parallelism to aggregate sparse performance measurements from every rank, thread, and GPU stream of an application, and attributes heterogeneous call path profiles and traces to source code.
Using the same amount of resources, our approach analyzes large-scale performance measurements of GPU-accelerated applications over an order of magnitude faster than HPCToolkit and its sparse analysis results are as much as three orders of magnitude smaller than HPCToolkit's dense representation of metrics.
\end{abstract}

\begin{document}

\maketitle

\section{Introduction}
Emerging exascale compute platforms pose significant challenges for performance tools.
%Performance tools must support measurement and analysis of application executions with potential performance problems that may only become apparent at very large scales.
Applications may have performance problems that only become apparent at very large scales, thus performance tools must  support  measurement and analysis at scale.
For applications running on tens of thousands of compute nodes equipped with multicore processors,\footnote{For example,
Riken's Fugaku supercomputer has 158,976 nodes, each equipped with 48 compute cores~\cite{fugaku}.} performance tools may record
gigabytes of measurement data per second.
Analysis of measurement data becomes more expensive as data volume grows.
%To make performance analysis feasible for exascale systems, tools must employ compact and effective storage formats along with efficient highly-parallel analysis algorithms.

Performance measurements for parallel programs are naturally sparse in two ways: the set of metrics present for any context and the set of contexts present in different threads. Metric sparsity can be very significant. For instance, on systems with GPU-accelerated compute nodes, such as the US DOE's emerging exascale supercomputers, metrics for GPU code are disjoint from those for CPU code. 
When both CPU and GPU metrics are collected for an application, 
%there is natural metric sparsity across code regions:
none of the myriad of GPU metrics apply to any of the many CPU contexts. Context sparsity may also be significant. Threads with different roles
typically execute different code in different calling contexts.
For tools to efficiently measure and analyze code performance on heterogeneous exascale systems, sparse metrics and contexts must be handled efficiently throughout tool workflows.

\begin{comment}
To make performance analysis feasible for large-scale executions on emerging GPU-accelerated exascale systems, tools must employ efficient highly-parallel, analysis algorithms to analyze large volumes of sparse measurement data using out-of-core methods.
\end{comment}
In this paper, we present an approach to address the aforementioned challenges: a novel, highly-parallel strategy for postmortem analysis that uses sparse formats and out-of-core methods to analyze performance data. Our approach is particularly effective for analyzing large-scale measurements of GPU-accelerated applications because of their extreme metric sparsity. We implement these methods in HPCToolkit-SA - a tool derived from the open-source HPCToolkit performance tools~\cite{GPUCtxReconstruction}.
HPCToolkit-SA aggregates call path profiles and traces from every CPU thread and GPU stream in the application and associates them with detailed heterogeneous calling contexts that include inlined functions, loops, and calls to device functions within GPU kernels. It does so in a way that is scalable and efficient in both time and space.
This paper makes the following contributions:

\begin{itemize}
\item it describes HPCToolkit-SA's novel sparse formats for measurement data and analysis results,
designed to accelerate key access patterns for postmortem analysis, presentation, and data analytics;
\item it outlines HPCToolkit-SA's novel, highly-parallel \textit{streaming aggregation} approach to postmortem analysis, which employs both shared and distributed memory parallelism, concurrent data structures, and efficient algorithms to rapidly analyze sparse measurements; and
\item it evaluates the effectiveness of HPCToolkit-SA's approach in terms of storage volume and analysis time.
\end{itemize}

% While our methods and software apply to analysis of performance measurements for both CPU and GPU-accelerated systems, they are exceptionally effective for analysis of measurements of GPU-accelerated applications because of extreme metric sparsity.
In our case studies analyzing measurement data for GPU-accelerated executions, HPCToolkit-SA's use of sparse formats instead of dense ones yielded as much as an order of magnitude reduction in the size of measurement data and three orders of magnitude reduction in the size of analysis results. 
Our analysis algorithm is capable of processing measurements from tens of thousands of application threads and GPU streams in minutes, using only a small fraction of the resources consumed by the application itself.
These results show the promise of our design for analyzing performance data from extreme scale executions.

The remainder of the paper is organized as follows.
\Cref{sec:related work} overviews approaches used by other performance tools in this space, with particular focus on scalablity.
\Cref{sec:approach} describes HPCToolkit-SA's novel approach to postmortem analysis based on sparse formats and a highly-parallel analysis algorithm.
\Cref{sec:experiments} provides an empirical analysis of HPCToolkit-SA's postmortem analysis workflow, comparing against similar open-source postmortem analysis tools.
\Cref{sec:conclusions} summarizes our conclusions.

\section{Related Work}\label{sec:related work}
Performance tools supporting GPU-accelerated applications differ widely in their approach to large-scale applications and in the level of detail they provide.

Some tools, such as NVIDIA's Nsight Systems~\cite{Nsight} and Extrae~\cite{ParaverExtrae}, rely on traces of an application's execution. 
For long-running applications traces become huge and thus extremely slow to analyze, which limits the scalability of these tools.
Although HPCToolkit-SA analyzes traces when present, our work focuses on the harder problem of aggregating a set of per-thread or per-GPU-stream call path profiles.

Some tools reduce the performance data volume by eliding details at scale, this includes Intel's VTune Profiler~\cite{intel-vtune}, ARM MAP~\cite{ARMMAP}, and ScalaTrace~\cite{ScalaTrace}.
VTune only provides summary statistics for MPI, I/O, GPU and CPU performance. MAP retains a limited (1000 by default) number of samples for each application thread or GPU stream. ScalaTrace compresses MPI traces by recognizing repetitive structure within and across traces.
Details of problems that arise at large scales may be elided by these approaches. Also, without detailed performance attribution, the causes behind issues become unnecessarily difficult to recover.
The work in this paper feasibly includes complete performance data for each application thread.

Tools based on the Score-P~\cite{ScoreP} measurement infrastructure exploit a single dimension of sparsity using the CUBE data format~\cite{CUBE}, this includes Scalasca~\cite{Scalasca}, TAU~\cite{tau}, and Vampir~\cite{vampir}.
In our experiments in \cref{sec:experiments:storage:case studies} we find that for GPU-accelerated programs two dimensions of performance data are highly sparse, exploiting only one is insufficient at scale.
Our sparse profile formats exploit both dimensions of sparsity for significant space reductions.

\begin{figure}[t]
  \centering
  \includegraphics[width=.47\textwidth]{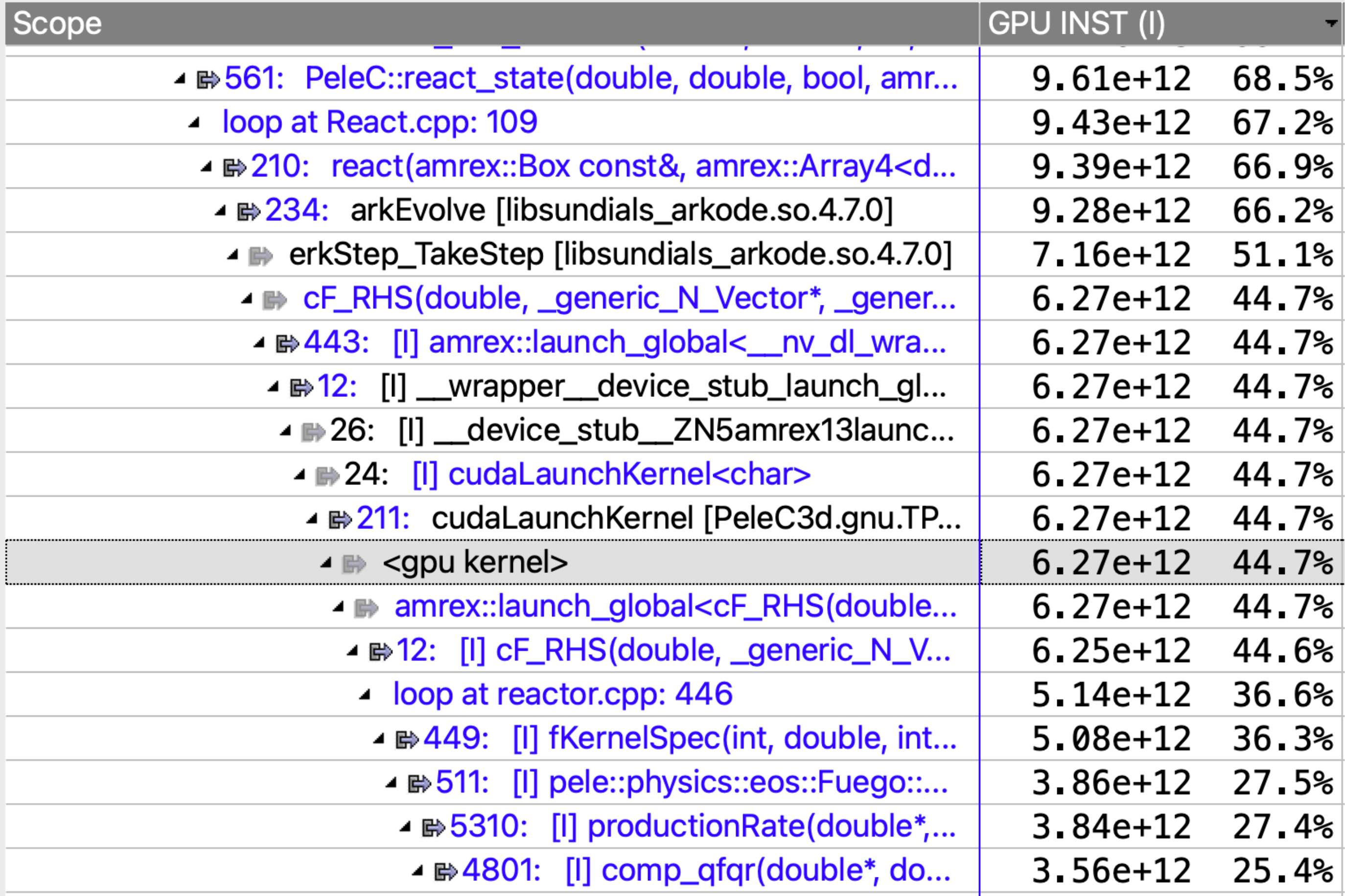}
  \caption{
    A segment of a heterogeneous call path containing functions, loops, and inlined code on both CPU and GPU.
  }
  \label{fig:hpctoolkit callstack}
\end{figure}

 HPCToolkit~\cite{GPUCtxReconstruction} attributes performance to complete heterogeneous calling contexts spanning both CPU and GPU code.
\Cref{fig:hpctoolkit callstack} shows the calling contexts surrounding a GPU kernel launch, lines above the highlighted line indicate the host-side launch site while lines below show a loop and inlined function calls within a device-side kernel.
Some tools only attribute GPU performance at the kernel-launch level, this includes Score-P~\cite{ScoreP}, Scalasca~\cite{Scalasca}, and TAU~\cite{tau}.
Other tools, such as Intel's VTune~\cite{intel-vtune} and NVIDIA's Nsight Compute~\cite{NCU} only support ``flat'' line-level attribution on the device side.
According to David Richards (LLNL), Mercury~\cite{mercury}--a next-generation, general-purpose radiation transport code--has O(100K) lines of code in a single kernel; the causes of performance losses in such a code  can't be fully understood without detailed attribution within an kernel.
HPCToolkit-SA supports detailed attribution of metrics within GPU kernels by using HPCToolkit's measurement infrastructure to collect performance data.

Finally, many tools do not use parallelism past a single compute node in their postmortem performance analysis phases, this includes NVIDIA's NSight Systems~\cite{Nsight}, Intel's VTune Profiler~\cite{intel-vtune}, TAU~\cite{tau} and Vampir~\cite{vampir}.
Extrae~\cite{ParaverExtrae} provides an MPI-based merger that integrates a collection of traces into a single Paraver trace file.
Scalasca's Scout~\cite{Scalasca} uses MPI+OpenMP parallelism to assemble profile results as dense blocks of metrics in the CUBE format.
HPCToolkit's {\tt hpcprof-mpi}~\cite{Toolkit:overview} uses pure-MPI parallelism to integrate multiple performance profiles into a dense tensor indexed by application thread, calling context, and metric.
Our highly-parallel analysis approach uses both thread-level and process-level parallelism to perform postmortem performance analysis, producing a compact sparse representation.

\section{Approach}\label{sec:approach}
To address the challenge of analyzing very large profiles (and optionally traces) for large-scale executions on either homogeneous (CPU) or heterogeneous (CPU+GPU) scalable parallel systems,
we created HPCToolkit-SA. HPCToolkit-SA differs from the open-source implementation of HPCToolkit for GPU-accelerated systems~\cite{GPUCtxReconstruction} in two key ways: it uses novel sparse representations for  performance data and it uses a novel \textit{streaming aggregation} approach for highly-parallel postmortem analysis of large, sparse performance measurement data.
\cref{sec:storage formats} describes the design of HPCToolkit-SA's sparse representations. \cref{sec:algorithm} presents HPCToolkit-SA's  streaming aggregation approach.

\subsection{Sparse Formats for Performance Data}\label{sec:storage formats}
%For any algorithm, a lower bound on the execution time can be computed by considering data volume to be read and written and the file system bandwidth.
%To analyze the large volumes of measurements produced by large-scale executions, the formats in which we store these measurements and their analysis results are an important consideration.
Applications running on tens of thousands of compute nodes can generate gigabytes of measurement data per second. 
Unsurprisingly, postmortem analysis of large measurement data also produces large analysis results.
For GPU-accelerated programs, metrics are numerous and very sparse. 
The sparsity appears in two ways: metrics and contexts.
Below, we list a few factors that contribute to metric and context sparsity for GPU-accelerated programs.
\begin{itemize}
\item
  Metrics accumulate within expensive leaf functions; often distant calling context ancestors representing high-level functions will have no exclusive metrics.
\item
  Many metrics apply only to instructions or code regions executed on a particular compute device, such as stall metrics for GPU instructions (12 kinds on NVIDIA GPUs) or CPU metrics for work, resource utilization, or waste.
\item
  Some metrics apply only to very specific contexts such as GPU kernels (e.g., GPU utilization, block count) or GPU data movement operations (e.g., copy costs in bytes and time).
\item
  \sloppy Threads with different roles (e.g., main thread, OpenMP worker thread, GPU progress thread, MPI I/O helper thread) typically execute different code in different calling contexts.
  After postmortem analysis, threads will have zero-valued metrics for contexts only present in other threads.
\end{itemize}

The number of metrics affected by these kinds of sparsity is large in practice, for instance our measurements of GPU-accelerated programs described later in \cref{sec:experiments} use a single CPU metric but 62 GPU metrics.
As part of the postmortem analysis, most of these exclusive metrics are augmented with inclusive metric costs and may be inputs to derived metrics.\footnote{
  For example, GPU utilization $= \frac{\rm measured\ GPU\ PC\ samples}{\rm expected\ GPU\ PC\ samples\ across\ all\ SMs}$.
}
None of the many GPU metrics apply to any CPU context, which are typically more numerous than GPU contexts.

We combat the increase in data volume by exploiting these two kinds of sparsity with a series of new sparse formats.
In addition to lower space requirements, data accesses need to be efficient to ensure the performance of postmortem analysis and responsiveness of a graphical browser for analysis results.
Therefore, a guarantee of efficient data accesses is a requirement of these new formats.

%Retaining empty values in measurement data slows postmortem analysis; retaining them in analysis results slows interactive exploration of results.
%To avoid the unnecessary cost of analyzing many zeros, we designed and implemented a series of sparse formats for both measurements data and analysis results.
%These formats are designed to reduce the total data volume and are optimized for streamlined access to related data entries.

%Our sparse formats were inspired by the Compressed Sparse Row (CSR) format for sparse matrices~\cite{CSR}.
%CSR reduces the data volume for a sparse matrix while providing fast access to individual values.
%An $m\times n$ sparse matrix with $x$ non-zero values in CSR format takes $\mathcal{O}(x+m)$ space instead of the $\mathcal{O}(mn)$ space required for a dense representation, and any individual value can be dereferenced in $\mathcal{O}(\log (x/m))$ time.

%To save storage and provide efficient access to non-zero metrics, we designed and implemented a series of sparse formats for both measurement data and analysis results. 
Our solution was inspired by the Compressed Sparse Row (CSR) format for sparse matrices~\cite{CSR}.
\Cref{sec:storage formats:measurement} describes our sparse format for measurement data.
\Cref{sec:storage formats:database} describes our two sparse formats for analysis results. %, arranged for efficient access of related analysis results.

\subsubsection{Sparse Measurement Format}\label{sec:storage formats:measurement}
\begin{figure}[t]
  \small
  \begin{tikzpicture}[>=stealth, label distance=-2pt]
    \matrix[matrix of math nodes, nodes in empty cells] (vcr) {
      |[label=left:{(m,v)\!:}]| \lbrack &[-2pt] (1, 5) & (0, 2) & (2, 6) & (1, 3) &[-2pt] \rbrack \\
      \\
      \\
      |[label=left:{(c,i)\!:}]| \lbrack & (0, 0) & (1, 1) & (3, 3) & (\top, 4) & \rbrack \\
    };
    \draw (vcr-4-2) edge[->] (vcr-1-2) (vcr-4-3) edge[->] (vcr-1-3)
          (vcr-4-4) edge[->] (vcr-1-5);
    \node[left=5mm of vcr] (dense) {$\begin{bmatrix}
      0 & 5 & 0 & 0 \\
      2 & 0 & 6 & 0 \\
      0 & 0 & 0 & 0 \\
      0 & 3 & 0 & 0 \\
    \end{bmatrix}$};
    \node[above=1mm of dense,anchor=center] {Metric $m$};
    \node[left=1mm of dense,anchor=center,rotate=90] {Context $c$};
    \draw (dense) edge[double,thick,->] (vcr);
  \end{tikzpicture}
  \caption{
    HPCToolkit-SA's sparse format for measurement data.
  }
  \label{fig:runfmt}
\end{figure}
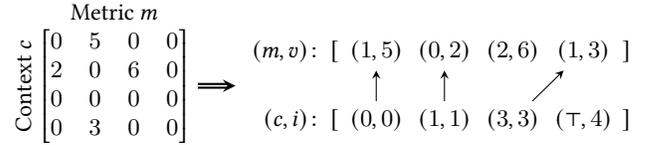

HPCToolkit profiles each CPU thread and GPU stream of an application, storing the per-thread profiles in separate measurement files.
Each measurement file includes a representation of the calling context tree observed in that thread or stream.
HPCToolkit stores metric values collected exclusively for each calling context in a dense vector stored as part of the calling context tree node.
In contrast, HPCToolkit-SA's measurement subsystem stores metric values in a separate region of a measurement file using the format shown in \cref{fig:runfmt} and described below.
%To better support the natural sparsity in the measurement data, we modified HPCToolkit to instead store the metric values in our sparse format, separately from the calling context tree .

The left side of \cref{fig:runfmt} shows a dense matrix storing metric values for all metrics and all calling contexts.
Each element of this matrix is the metric value $v$ measured for a metric $m$ in a calling context $c$.
Our sparse format is shown on the right, consisting of an $(m,v)$ vector listing non-zero metric values and a $(c,i)$ vector mapping contexts to contiguous ranges of the $(m,v)$ vector\footnote{
  $i$ is the index of the first $(m,v)$ associated with a calling context $c$.
  The end of $c$'s range directly precedes the following $i$ or ending $\top$ sentinel.
}.
The $(c,i)$ vector is sorted by $c$ and the $(m,v)$ vector by $m$ within each contiguous context range.
These two sorts enable the use of binary search to access the data for an individual context or metric.

This format takes advantage of both kinds of sparsity present in this measurement matrix.
First, ``empty'' calling contexts with all zero metric values are elided from the $(c,i)$ vector.
%First, ``empty'' calling contexts with no associated non-zero metric values can be elided from the $(c,i)$ vector, thus taking no additional space.
%These contexts may be caused by inner functions which only call other more expensive leaf functions, and thus are never sampled at the top of the execution stack.
Second, zeros in the metric values for any context are elided from the $(m,v)$ vector.
%Second, zero metric values are elided from the $(m,v)$ vector, thus many metrics may be zero for a calling context at no additional space cost.
%One cause for this is the presence of enabled GPU metrics, which are always zero in CPU calling contexts.
In short, we exploit both row-wise and column-wise sparsity within this matrix.

Let  $\mathcal{V}$, $\mathcal{M}$ and $\mathcal{C}$ be the number of non-zero metric values, metrics and non-empty calling contexts respectively.
The space cost of our sparse format is $\mathcal{O}(\mathcal{V}+\mathcal{C})$.
%The space cost of our sparse format is equivalent to that of CSR, $\mathcal{O}(\mathcal{V}+\mathcal{C})$.
%We also sort the $(m,v)$ and $(c,i)$ vectors by $m$ and $c$ respectively, thus 
In $(m,v)$, referencing the start of the range for a calling context $c$ can be done in $\mathcal{O}(\log \mathcal{C})$ time with a binary search, while accessing a specific value takes another $\mathcal{O}(\log \mathcal{M})$ time binary search within the mapped range.
%We also sort the $(m,v)$ and $(c,i)$ vectors by $m$ and $c$ respectively to recover the logrithmic $\mathcal{O}(\log (\mathcal{V}/\mathcal{C}))$ access time.
%Although not required for iteration, this sorting allows us to reuse this pattern in our database format as discussed in the following section.

\subsubsection{Sparse Analysis Result Formats}\label{sec:storage formats:database}
A browser interactively exploring performance analysis results requires efficient access to metrics both within a single profile and for selected contexts across multiple profiles.
For instance, a developer may investigate the performance of an individual CPU thread across all calling contexts, or check for load imbalance across threads within a particular calling context.
Today, HPCToolkit outputs one dense result file for each CPU thread and each GPU stream.
This is acceptable for the first scenario but not the second, as a browser would need to read a single value from each of thousands of files, each of which may be quite large.

\begin{figure}[t]
  \centering
  \small
  \begin{tikzpicture}[scale=.8]
    \draw[thick] (0,0) -- +(.1,-.1)
          +(.2,-.2) coordinate (cm 4)
          +(0,0) -- coordinate[pos=.4] (cm 3 split) coordinate[pos=.68] (cm 2 split)
            node[auto,sloped] {Profile $p$}
          ++(40:2.5) -- +(.1,-.1)
          +(.2,-.2) coordinate (cm 1);
    \path (cm 2 split) +(.25,-.25) node[rotate=40] {\large$\cdots$} (cm 3 split) +(.2,-.2) coordinate (cm 3);
    \foreach \coord in {cm 1,cm 3,cm 4} {
      \draw[fill=black!0!white,thick] (\coord) rectangle +(2.6,-2.6);
      \foreach \lenx in {0,.2,...,2.4} {
        \foreach \leny in {0,.2,...,2.4} {
          \pgfmathparse{(rnd > 0.8) ? 0.2 : 0} \edef\x{\pgfmathresult}
          \path[fill=black!30!white] (\coord) ++(\lenx,-\leny) rectangle +(\x,-\x);
        }
      }
      \foreach \len in {.2,.4,...,2.4} {
        \draw[color=black!40!white]
          (\coord) ++(\len,0) -- +(0,-2.6)
          (\coord) ++(0,-\len) -- +(2.6,0);
      }
      \draw[draw=black,thick] (\coord) rectangle coordinate[midway] (front center) +(2.6,-2.6);
    }
    \path (cm 1) -- node[auto] {Metric $m$} +(2.6,0);
    \path (cm 4) ++(0,-2.6) -- node[auto,sloped] {Context $c$} +(0,2.6);
    \node[fill=black!0!white, align=right, inner sep=2pt] (front arrays) at (front center) {
      $(m,v)\!: \lbrack\hspace{2em}\rbrack$ \\
      $\uparrow \uparrow \dots$ \\
      $(c,i)\!: \lbrack\hspace{2em}\rbrack$
    };
    \node[below=12mm of front center] (p array) {
      %$(p,j)\!: \lbrack\, (p, 0) \dots \,\rbrack$
      $p\!: \lbrack\, 0 \dots \,\rbrack$
    };
    \draw[->,thick] ($(p array.north) + (0em,-.7mm)$) -- ($(front arrays.south) + (.6em, .7mm)$);

    \draw[thick] (5.5,0) -- +(.1,-.1)
          +(.2,-.2) coordinate (cm 4)
          +(0,0) -- coordinate[pos=.4] (cm 3 split) coordinate[pos=.68] (cm 2 split)
            node[auto,sloped] {Context $c$}
          ++(40:2.5) -- +(.1,-.1)
          +(.2,-.2) coordinate (cm 1);
    \path (cm 2 split) +(.25,-.25) node[rotate=40] {\large$\cdots$} (cm 3 split) +(.2,-.2) coordinate (cm 3);
    \foreach \coord in {cm 1,cm 3,cm 4} {
      \draw[fill=black!0!white,thick] (\coord) rectangle +(2.6,-2.6);
      \foreach \lenx in {0,.2,...,2.4} {
        \foreach \leny in {0,.2,...,2.4} {
          \pgfmathparse{(rnd > 0.8) ? 0.2 : 0} \edef\x{\pgfmathresult}
          \path[fill=black!30!white] (\coord) ++(\lenx,-\leny) rectangle +(\x,-\x);
        }
      }
      \foreach \len in {.2,.4,...,2.4} {
        \draw[color=black!40!white]
          (\coord) ++(\len,0) -- +(0,-2.6)
          (\coord) ++(0,-\len) -- +(2.6,0);
      }
      \draw[draw=black,thick] (\coord) rectangle coordinate[midway] (front center) +(2.6,-2.6);
    }
    \path (cm 1) -- node[auto] {Profile $p$} +(2.6,0);
    \path (cm 4) ++(0,-2.6) -- node[auto,sloped] {Metric $m$} +(0,2.6);
    \node[fill=black!0!white, align=right] (front arrays) at (front center) {
      $(p,v)\!: \lbrack\hspace{2em}\rbrack$ \\
      $\uparrow \uparrow \dots$ \\
      $(m,i)\!:\lbrack\hspace{2em}\rbrack$
    };
    \node[below=12mm of front center] (p array) {
      %$(c,j)\!: \lbrack\, (c, 0) \dots \,\rbrack$
      $c\!: \lbrack\, 0 \dots \,\rbrack$
    };
    \draw[->,thick] ($(p array.north) + (0em,-.7mm)$) -- ($(front arrays.south) + (.6em, .7mm)$);
  \end{tikzpicture}
  \caption{
    Profile Major Sparse (left) and Context Major Sparse (right) analysis result storage formats.
  }
  \label{fig:sf}
\end{figure}
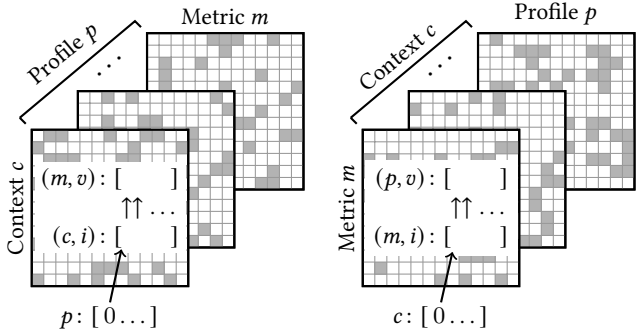

%A browser interactively exploring performance analysis results requires efficient access to metrics both within a single profile and for selected contexts across multiple profiles.
%For instance, a developer may investigate the performance of an individual CPU thread across all calling contexts, or check for load imbalance across threads within a particular calling context.
To support efficient access for both use cases while reducing space requirements, for HPCToolkit-SA we developed two formats for these two use cases, respectively: \textit{Profile Major Sparse} (PMS) format and \textit{Context Major Sparse} (CMS) format.
Each format contains the entirety of the analysis results in a single file to avoid large numbers of files. Their main difference is in the order in which the performance data is stored, allowing readers to read the file best suited for their use case.
\Cref{fig:sf} shows a high-level visualization of these formats. %, composed of multiple planes arranged in the format discussed in the previous section. 
For brevity we only detail the PMS format, CMS is similar with dimensions transposed as shown in \cref{fig:sf}.
%Unlike measurement data which can be represented as a 2-dimensional matrix, analysis results are 3-dimensional with the addition of a profile $p$ dimension, indicating the related CPU thread or GPU stream.

%For brevity we detail only the PMS format, CMS is similar with dimensions transposed as shown in \cref{fig:sf}.
Our PMS format consists of $\mathcal{P}$ ``planes'' of data each representing the analysis results for a single profile $p$. Each plane contains results for every calling context $c$ and metric $m$ for that profile in the CSR-like format shown in \cref{fig:runfmt}, which exploits both metric and context sparsity.
Planes are referenced by byte offset in a $\mathcal{P}$-sized ``table of contents'' header, allowing planes to appear in the file in any order.
Since every profile generally has at least one non-zero analysis result, we do not need to use a sparse representation for this mapping.

Let $\mathcal{V}$, $\mathcal{M}$ and $\mathcal{C}$ be the number of non-zero values, metrics and non-empty calling contexts respectively, averaged across all $\mathcal{P}$ profiles.
The space cost of our PMS format is $\mathcal{O}(\mathcal{P}(\mathcal{V}+\mathcal{C}))$.
The plane for any particular profile can be obtained from the ``table of contents'' in $\mathcal{O}(1)$ time. Then as noted before, the range for a particular context can be referenced in $\mathcal{O}(\log \mathcal{C})$ time and any specific value in an additional $\mathcal{O}(\log \mathcal{M})$ time using binary search.
Similar algorithms with transposed dimensions can be used in the CMS format to access the data for an individual context and metric in logarithmic time.
In total, our PMS and CMS formats provide efficient access to all analysis results while exploiting sparsity to compress the contained performance data.

\subsection{Streaming Aggregation}\label{sec:algorithm}
We require an efficient approach to correlate and aggregate data measured from thousands of application threads or GPU streams.
We divide the postmortem analysis performed by HPCToolkit-SA into three classes based on common dataflow properties: class 1 analyses are produced for each performance profile of a single CPU thread or GPU stream, while class 2 analyses correlate and class 3 analyses aggregate data from multiple profiles.
Specifically, HPCToolkit-SA produces the following analyses:
\begin{itemize}
\item (Class 1) Inclusive and exclusive cost metrics for every calling context within a single thread or stream,
\item (Class 1) Execution traces of each thread or stream, composed of a sequence of samples containing a timestamp and calling context,
\item (Class 2) Unified source-level calling context tree, in which each node represents a unique call path observed at runtime,
\item (Class 2) Unified source files and application binaries, in which each node represents a unique source file or binary used by the application, and
\item (Class 3) Inclusive and exclusive cost statistics for every calling context and metric, summarizing the full application execution.
\end{itemize}

To produce these analyses, HPCToolkit-SA employs a novel thread-parallel approach we call \textit{streaming aggregation} to process performance data, described in \cref{sec:algorithm:threads}.
This approach provides excellent performance and makes use of shared-memory algorithms to provide better efficiency than a purely MPI-based approach.
For scalability, HPCToolkit-SA also uses conventional distributed-memory parallelism across multiple compute nodes in a hybrid approach, described in \cref{sec:algorithm:mpi}.
Finally, in \cref{sec:algorithm:lexical,sec:algorithm:output} we detail the parallelizations we chose for a selection of the algorithms required for producing our analysis.

\subsubsection{Thread-level Parallelism}\label{sec:algorithm:threads}
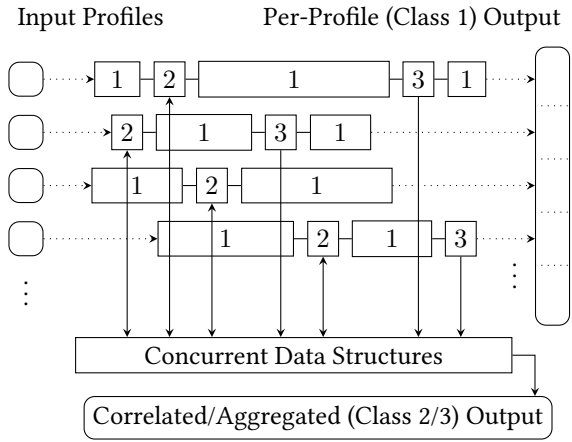
\begin{figure}[t]
  \centering
  \begin{tikzpicture}[>=stealth]
    \node[draw,rounded corners,minimum height=3.7cm,minimum width=3ex] (profile output) at (7cm,-1.8cm) {};
    \node[above left=.5ex and 0pt of profile output.north east] (profile output label) {Per-Profile (Class 1) Output};

    \node[draw,rounded corners,minimum size=3ex] (profile 1) at (0,-2.5ex) {};
    \node[draw,rounded corners,minimum size=3ex,below=1.6ex of profile 1] (profile 2) {};
    \node[draw,rounded corners,minimum size=3ex,below=1.6ex of profile 2] (profile 3) {};
    \node[draw,rounded corners,minimum size=3ex,below=1.6ex of profile 3] (profile 4) {};
    \coordinate (profiles anchor) at ($(profile 1.west)!.5!(profile 4.west)$);
    \node[anchor=south west] at (profile output label.south -| profile 1.west) {Input Profiles};

    \coordinate (profile midpoint 1 2) at ($(profile 1)!.5!(profile 2)$);
    \coordinate (profile midpoint 2 3) at ($(profile 2)!.5!(profile 3)$);
    \coordinate (profile midpoint 3 4) at ($(profile 3)!.5!(profile 4)$);
    \coordinate (profile midpoint 4 n) at ($(profile 3)!1.5!(profile 4)$);
    \draw[dotted]
      (profile midpoint 1 2 -| profile output.west) -- (profile midpoint 1 2 -| profile output.east)
      (profile midpoint 2 3 -| profile output.west) -- (profile midpoint 2 3 -| profile output.east)
      (profile midpoint 3 4 -| profile output.west) -- (profile midpoint 3 4 -| profile output.east)
      (profile midpoint 4 n -| profile output.west) -- (profile midpoint 4 n -| profile output.east);

    \node[draw,right=.9cm of profile 2] (b1) {$2$};
    \coordinate (b2 anchor) at (profile 1 -| b1);
    \node[draw,right=1em of b2 anchor] (b2) {$2$};
    \coordinate (b3 anchor) at (profile 3 -| b2);
    \node[draw,right=1em of b3 anchor] (b3) {$2$};
    \coordinate (b4 anchor) at (profile 2 -| b3);
    \node[draw,right=2em of b4 anchor] (b4) {$3$};
    \coordinate (b5 anchor) at (profile 4 -| b4);
    \node[draw,right=1em of b5 anchor] (b5) {$2$};
    \coordinate (b6 anchor) at (profile 1 -| b5);
    \node[draw,right=3em of b6 anchor] (b6) {$3$};
    \coordinate (b7 anchor) at (profile 4 -| b6);
    \node[draw,right=1em of b7 anchor] (b7) {$3$};

    \node (a1 text) at ($(b1)!.5!(b4)$) {$1$};
    \node[draw,inner sep=0,fit={(a1 text) ($(b1.east)+(.5em,0)$) ($(b4.west)-(.5em,0)$)}] (a3) {};
    \node (a2 text) at ($(b2)!.5!(b6)$) {$1$};
    \node[draw,inner sep=0,fit={(a2 text) ($(b2.east)+(.5em,0)$) ($(b6.west)-(.5em,0)$)}] (a5) {};
    \node (a3 text) at ($(b5)!.5!(b7)$) {$1$};
    \node[draw,inner sep=0,fit={(a3 text) ($(b5.east)+(.5em,0)$) ($(b7.west)-(.5em,0)$)}] (a8) {};

    \node[draw,left=.5em of b2,minimum width=.6cm] (a1) {$1$};
    \node[draw,left=.5em of b3,minimum width=1.2cm] (a2) {$1$};
    \node[draw,left=.5em of b5,minimum width=1.8cm] (a4) {$1$};
    \node[draw,right=.5em of b6,minimum width=.5cm] (a9) {$1$};
    \node[draw,right=.5em of b4,minimum width=.8cm] (a7) {$1$};
    \node[draw,right=.5em of b3,minimum width=2cm] (a6) {$1$};

    \draw[->,dotted]
      (profile 1) edge (a1)
      (profile 2) edge (b1)
      (profile 3) edge (a2)
      (profile 4) edge (a4);
    \draw[->,dotted]
      (a9) edge (a9 -| profile output.west)
      (a7) edge (a7 -| profile output.west)
      (a6) edge (a6 -| profile output.west)
      (b7) edge (b7 -| profile output.west);
    \draw
      (a1) -- (b2) -- (a5) -- (b6) -- (a9)
      (b1) -- (a3) -- (b4) -- (a7)
      (a2) -- (b3) -- (a6)
      (a4) -- (b5) -- (a8) -- (b7);

    \node[below=0pt of profile 4] (left dots) {$\vdots$};
    \coordinate (right dots head) at ($(b7.east)!.5!(b7.east -| profile output.south)$);
    \node[below=0pt of right dots head] (right dots) {$\vdots$};

    \node[draw,below left=1ex and .3cm of profile output.south west,minimum width=5.8cm] (data structures) {Concurrent Data Structures};
    \node[draw,rounded corners,anchor=north east] (aggregate output) at ($(profile output.north |- data structures.south)+(0,-2ex)$)
        {Correlated/Aggregated (Class 2/3) Output};
    \draw[->] (data structures.east) -| (aggregate output.north -| profile output.west);

    \draw[<->]
      (b1) edge (data structures.north -| b1)
      (b2) edge (data structures.north -| b2)
      (b3) edge (data structures.north -| b3)
      (b5) edge (data structures.north -| b5);
    \draw[->]
      (b4) edge (data structures.north -| b4)
      (b6) edge (data structures.north -| b6)
      (b7) edge (data structures.north -| b7);
  \end{tikzpicture}
  \caption{
    Conceptual diagram of the streaming aggregation thread-parallel approach, discussed in \cref{sec:algorithm:threads}.
    Numbered blocks indicate operations of the numbered class, as defined in \cref{sec:algorithm}.
  }
  \label{fig:streaming aggregation}
\end{figure}

To analyze performance data from large-scale executions, our thread-parallel streaming aggregation approach exploits formal properties of the three classes of analyses HPCToolkit-SA produces.
First, since class 1 analyses are produced for every input performance profile, much of their processing is independent of other profiles and can be performed in parallel.
Second, class 2 and 3 analyses do not enforce any particular order, they are associative, commutative, and class 2 analyses are also idempotent.

\Cref{fig:streaming aggregation} shows the conceptual layout of our streaming aggregation approach.
Class 1 analyses are performed using conventional embarrassing parallelism, each thread reads and processes data from an input profile and writes the class 1 analysis results directly to the output before moving on to the next profile.
This design means class 1 data does not reside in memory for the lifetime of the analysis, in our approach class 1 data effectively ``streams'' from the input profiles to the output analysis results.

This design also means that producing class 2 and 3 analyses requires communication across the threads.
Since class 2/3 analyses can be freely reordered, we implemented this communication as a series of shared concurrent data structures where the result of every operation integrates a class 2/3 datum.
This design means only the class 2/3 analysis results reside in memory, ``aggregated'' from the data parsed from each input profile.
Conventionally this aggregation would be done using reduction phases, our decision to use concurrent data structures was informed by the preliminary experiment described later in \cref{sec:lesson:reductions are slow}, which finds that this use of shared concurrent data structures significantly outperforms conventional reduction-based approaches.

Class 2 analyses differ from class 3 analyses in that class 2 analyses are required to output class 1 analyses, while class 1 analyses are aggregated to form class 3 analyses.
An important class 2 example is calling context identifiers, which must be unique to each calling context but consistent across profiles.
To support these cases we perform class 2 analyses as early as possible, for instance we assign identifiers exactly when a new, unique calling context has been added to the shared data structures.
These class 2 analyses are not allowed to change afterward, thus providing the consistency across profiles while retaining the high degree of parallelism afforded by our thread-parallel approach.

\subsubsection{Process-level Parallelism}\label{sec:algorithm:mpi}
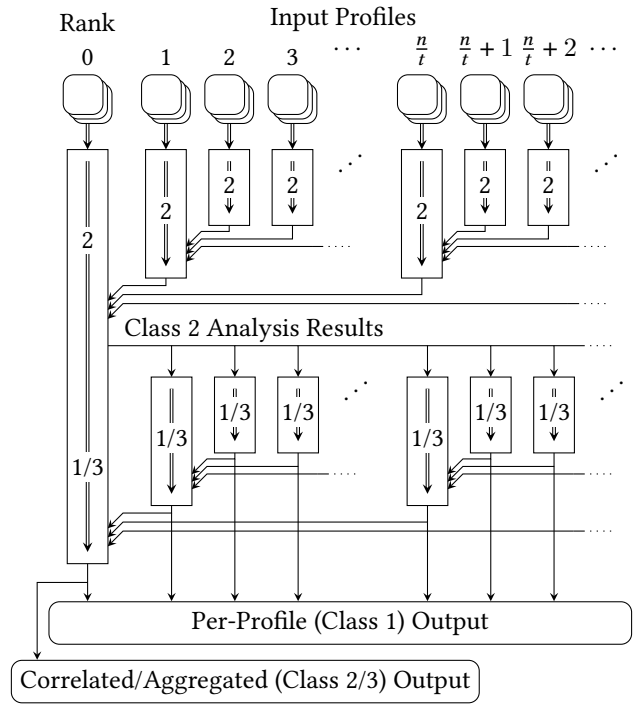
\begin{figure}[t]
  \centering
  \begin{tikzpicture}[>=stealth]
    \coordinate ({origin} src) at (0,0);
    \foreach \line/\sep/\height/\rank [remember=\line as \lastline (initially origin)] in {
      master/0pt/52mm/0,
      worker0 top/5mm/14mm/1, worker00 top/3mm/7mm/2, worker01 top/3mm/7mm/3, worker0 top dots/3mm/0pt/{},
      worker1 top/5mm/14mm/{$\frac{n}{t}$}, worker10 top/3mm/7mm/{$\frac{n}{t}+1$},
      worker11 top/3mm/7mm/{$\frac{n}{t}+2$}, worker1 top dots/3mm/0pt/{}} {
      \IfSubStr{\line}{dots}{
        \node[inner sep=0pt,right=\sep of {\lastline} src] ({\line} src) {$\iddots$};
        \node[above=1cm of {\line} src] { $\dots$};
      }{
        \node[inner sep=2pt,right=\sep of {\lastline} src, minimum width=1.5em] ({\line} src) {};
        \node[above=1cm of {\line} src] {\rank};
        \node[inner sep=2pt,below=\height of {\line} src, minimum width=1.5em] ({\line} sink) {};
        \draw[->,double] ({\line} src) -- coordinate[midway] ({\line} wavemark) ({\line} sink);
        \coordinate ({\line} wavemark) at ($({\line} wavemark)+(0,.4mm)$);
        \IfSubStr{\line}{master}{
          \node[inner sep=2pt,fill=white] at ($({\line} wavemark)+(0,10ex)$) {2};
          \node[inner sep=2pt,fill=white] at ($({\line} wavemark)-(0,10ex)$) {1/3};
        }{
          \node[inner sep=2pt,fill=white] at ({\line} wavemark) {2};
        }
        \draw ({\line} sink.west |- {\line} src.north) rectangle ({\line} sink.south east);
      }
    }
    \node[above=1cm+3ex of {master} src] (rank label) {Rank};

    \path ({master} src.west); \pgfgetlastxy{\Xwest}{\Ydummy}
    \path ({worker1 top dots} src.east); \pgfgetlastxy{\Xeast}{\Ydummy}
    \node[below left=5mm and 2.5mm of {master} sink,anchor=north west,draw,rounded corners,
          minimum width=4mm+\Xeast-\Xwest]
      (db) {Per-Profile (Class 1) Output};
    \coordinate (xmid) at ($({master} src.west)!.5!({worker1 top dots} src.east)$);
    \node[anchor=south] at (xmid |- rank label.south) {Input Profiles};
    \foreach \line in {master, worker0 top, worker00 top, worker01 top, worker1 top, worker10 top, worker11 top} {
      \foreach \o [count=\i] in {-.6mm, 0mm, .6mm} {
        \node[above=4mm of {\line} src,,draw,rounded corners,fill=white,minimum size=3.5ex,
              xshift=-\o,yshift=\o]
          ({\line} hpcrun {\i}) {};
      }
      \draw[->,double] ({\line} hpcrun {1}.south -| {\line} src) -- ({\line} src.north);
    }
    \draw[->] ({master} sink.south) -- coordinate (branch) ({master} sink.south |- db.north);
    \node[below left=2mm and 5mm of db,anchor=north west,draw,rounded corners] (exml) {Correlated/Aggregated (Class 2/3) Output};
    \path (exml.west |- db.west) -- coordinate[pos=.7] (branch2) (db.west);
    \draw[->] (branch) -| (branch2) -- (branch2 |- exml.north);

    \coordinate ({master bot} src) at ({worker0 top} sink.south -| {master} src.center);
    \node[inner sep=2pt,below=12mm of {master bot} src] ({master bot} src) {\phantom{ Src}};
    \path ({master bot} src.north -| {master} sink.east) ++(0,3mm) coordinate (bcast edge);
    \foreach \line/\sep/\height [remember=\line as \lastline (initially master bot)] in {
      worker0 bot/5mm/14mm,worker00 bot/3mm/7mm,worker01 bot/3mm/7mm,worker0 bot dots/3mm/0pt,
      worker1 bot/5mm/14mm,worker10 bot/3mm/7mm,worker11 bot/3mm/7mm,worker1 bot dots/3mm/0pt} {
      \IfSubStr{\line}{dots}{
        \node[inner sep=0pt,right=\sep of {\lastline} src] ({\line} src) {$\iddots$};
      }{
        \node[inner sep=2pt,right=\sep of {\lastline} src, minimum width=1.5em] ({\line} src) {};
        \draw[->] (bcast edge) -| coordinate[midway] (bcast edge) ({\line} src.north);
        \node[inner sep=2pt,below=\height of {\line} src, minimum width=1.5em] ({\line} sink) {};
        \draw[->,double] ({\line} src) -- coordinate[midway] ({\line} wavemark) ({\line} sink);
        \coordinate ({\line} wavemark) at ($({\line} wavemark)+(0,.4mm)$);
        \node[inner sep=2pt,fill=white] at ({\line} wavemark) {1/3};
        \draw ({\line} sink.west |- {\line} src.north) rectangle ({\line} sink.south east);
      }
    }
    \draw[dash pattern=on .5pt off 2pt on .5pt off 2pt on .5pt off 2pt on .5pt off 2pt on 2cm]
        (bcast edge -| {worker1 bot dots} src.center) -- (bcast edge);
    \node[anchor=south west,inner sep=1pt,xshift=.5em] at (bcast edge -| {master} sink.east) {Class 2 Analysis Results};
    \foreach \line/\o/\parentline [remember=\line as \lastline] in {
      worker00 top/.8mm/worker0 top, worker01 top/1.8mm/worker0 top, worker0 top dots/2.8mm/worker0 top,
      worker10 top/.8mm/worker1 top, worker11 top/1.8mm/worker1 top, worker1 top dots/2.8mm/worker1 top,
      worker0 top/1mm/master, worker1 top/2.2mm/master, worker1 top dots/3.4mm/master,
      worker00 bot/.8mm/worker0 bot, worker01 bot/1.8mm/worker0 bot, worker0 bot dots/2.8mm/worker0 bot,
      worker10 bot/.8mm/worker1 bot, worker11 bot/1.8mm/worker1 bot, worker1 bot dots/2.8mm/worker1 bot,
      worker0 bot/1mm/master, worker1 bot/2.2mm/master, worker1 bot dots/3.4mm/master} {
      \IfSubStr{\line}{dots}{
        \coordinate (start) at ($({\lastline} sink.south)-(0,\o)$);
        \coordinate (start) at (start -| {\line} src.center);
        \coordinate (target) at (start -| {\parentline} sink.east);
        \draw[dash pattern=on .5pt off 2pt on .5pt off 2pt on .5pt off 2pt on .5pt off 2pt on 2cm] (start) -- ++(-5mm,0) coordinate (start);
      } {
        \coordinate (start) at ({\line} sink.south);
        \coordinate (target) at ($({\line} sink.south)-(0,\o)$);
        \coordinate (target) at (target -| {\parentline} sink.east);
      }
      \coordinate (corner) at ($(target) + (2mm,0)$);
      \coordinate (end) at ($(corner)!1.44!45:(target)$);
      \draw[->] (start) |- (corner) -- (end);
      \IfSubStr{\line}{bot}{
        \IfSubStr{\line}{dots}{}{
          \draw[->] (start |- corner) coordinate (tmp) -- (tmp |- db.north);
        }
      }{}
    }
  \end{tikzpicture}
  \caption{
    Conceptual diagram of our process-parallel approach, discussed in \cref{sec:algorithm:mpi}.
    Each of the $\boldsymbol{n}$ ranks use $\boldsymbol{t}$ threads.
    Numbered blocks indicate instances of streaming aggregation limited to the numbered classes of operations.
  }
  \label{fig:streaming aggregation mpi extension}
\end{figure}

To exploit multiple compute nodes for portmortem analysis, we use more conventional distributed-memory parallelism implemented with MPI.
Each MPI rank in this approach uses our thread-parallel streaming aggregation approach to efficiently use the threads available on each compute node.

This hybrid MPI+SA approach is depicted conceptually in \cref{fig:streaming aggregation mpi extension}.
Since class 2 analysis is required for correct class 1 output from each rank, we reduce and distribute class 2 analysis results prior to performing class 1 analysis on any rank.
A similar reduction is done at the end of execution to aggregate class 3 data.
Class 1 data is never communicated across ranks, instead every rank writes out class 1 data to disk in parallel as discussed in \cref{sec:algorithm:threads}.
This allows our hybrid MPI+SA approach to retain the same ``streaming'' qualities of standard streaming aggregation, using MPI communication only during the ``aggregation'' across multiple ranks.

Both reductions in our hybrid approach are implemented using conventional reduction trees, however unlike a pure-MPI reduction tree we process the reduction inputs in parallel at every rank, using the thread-level parallelism afforded by  streaming aggregation.
In this case a $t$-ary reduction tree has minimal depth and an optimally short critical path, where $t$ is the number of threads available in each rank.
Thus for our hybrid approach we use $t$-ary reduction trees for the two reductions, which reduces to a single-round reduction when $t$ is larger than the number of ranks.
By using only two reductions in our MPI+SA hybrid approach, we retain much of the parallelism provided by our base streaming aggregation approach.

\subsubsection{Generating Source-level Calling Context}\label{sec:algorithm:lexical}
\begin{figure}[t]
  \centering
  \begin{tikzpicture}[>=stealth, anchor=west, parent anchor=south west,
      edge from parent path={(\tikzparentnode\tikzparentanchor) ++(.3,0) |- (\tikzchildnode\tikzchildanchor)},
      grow via three points={one child at (0,-.4) and two children at (0,-.4) and (0,-.8)}]
    \scriptsize
    \node (orig 10) {\texttt{+0x10}}
      child[missing] child[missing]
      child {node (orig 20) {\texttt{+0x20}} }
      child[missing]
      child {node (orig 28) {\texttt{+0x28}}
        child[missing] child[missing]
        child {node (orig 28 80) {\texttt{+0x80}} }
      }
      child[missing] child[missing] child[missing]
      child {node (orig 2b) {\texttt{+0x2B}}
        child[missing] child[missing]
        child {node (orig 2b 80) {\texttt{+0x80}} }
      }
      child[missing] child[missing] child[missing] child[missing] child[missing]
      child {node (orig 30) {\texttt{+0x30}} }
      child {node (orig 38) {\texttt{+0x38}} }
      child[missing] child[missing] child[missing]
      child {node (orig 3b) {\texttt{+0x3B}} }
      child {node (orig 00) {\texttt{+0x00}} }
    ;
    \node[above right=.64cm and 1cm of orig 10,anchor=west] (full)
      {\texttt{main()}}
      child {node {\texttt{main.c:5}}
        child[xshift=-1mm] {node (full 10) {\texttt{+0x10}}
          child {node {\texttt{foo()}}
            child {node {\texttt{foo.c:5}}
              child[xshift=-1mm] {node (full 20) {\texttt{+0x20}}}
            }
            child[missing]
            child {node {\texttt{foo.c:6}}
              child[xshift=-1mm] {node (full 28) {\texttt{+0x28}}
                child {node {\texttt{bar()}}
                  child {node {\texttt{bar.c:5}}
                    child[xshift=-1mm] {node (full 28 80) {\texttt{+0x80}} }
                  }
                }
              }
              child[missing] child[missing] child[missing]
              child[xshift=-1mm] {node (full 2b) {\texttt{+0x2B}}
                child {node {\texttt{bar()}}
                  child {node {\texttt{bar.c:5}}
                    child[xshift=-1mm] {node (full 2b 80) {\texttt{+0x80}} }
                  }
                }
              }
            }
            child[missing] child[missing] child[missing] child[missing]
            child[missing] child[missing] child[missing] child[missing]
            child {node {\texttt{loop at foo.c:7}}
              child[xshift=-5mm] {node {\texttt{foo.c:8}}
                child[xshift=-1mm] {node (full 30) {\texttt{+0x30}} }
                child[xshift=-1mm] {node (full 38) {\texttt{+0x38}} }
              }
            }
            child[missing] child[missing] child[missing]
            child {node {\texttt{foo.c:9}}
              child[xshift=-1mm] {node {\texttt{(inlined) baz()}}
                child[xshift=-5mm] {node {\texttt{foo.c:20}}
                  child[xshift=-1mm] {node (full 3b) {\texttt{+0x3B}} }
                }
              }
            }
          }
          child[missing] child[missing] child[missing] child[missing] child[missing]
          child[missing] child[missing] child[missing] child[missing] child[missing]
          child[missing] child[missing] child[missing] child[missing] child[missing]
          child[missing] child[missing] child[missing] child[missing]
          child {node (full 00) {\texttt{+0x00}} }
        }
      }
    ;
    \draw[dashed]
      (orig 10) edge (full 10)
      (orig 20) edge (full 20)
      (orig 28) edge (full 28)
      (orig 28 80) edge (full 28 80)
      (orig 2b) edge (full 2b)
      (orig 2b 80) edge (full 2b 80)
      (orig 30) edge (full 30)
      (orig 38) edge (full 38)
      (orig 3b) edge (full 3b)
      (orig 00) edge (full 00)
    ;
  \end{tikzpicture}
  \caption{
    Expansion of calling context recorded at runtime with lexical information from offline analysis of application binaries by HPCToolkit's {\tt hpcstruct}.
  }
  \label{fig:lexical expansion}
\end{figure}
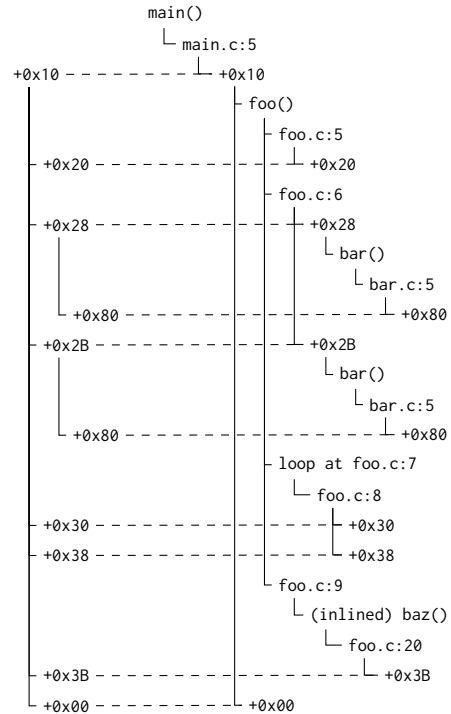

HPCToolkit's measurement infrastructure records observed calling contexts as the addresses of individual instructions encountered during a call stack unwind, specifically as byte offsets within application binaries.
To provide useful performance analysis, HPCToolkit-SA expands these instruction-level contexts into source-level (lexical) calling context, including functions, inlined function calls, nested loops and source lines, as shown in \cref{fig:lexical expansion}.
For simplicity, we perform this expansion for every context read for a profile, creating only the expanded sequence of contexts shown on the right of the figure in the shared context tree.

In this design, each context present in a profile requires a lookup in the source-level mappings available for an application binary, which can be obtained from DWARF debugging information or pre-calculated with HPCToolkit's {\tt hpcstruct} tool.
We parse both of these formats serially, which for large binaries may cause unwanted serialization.
To combat this, we aggregate application binaries from profiles first, and eagerly load source mappings for every new binary found.
After the first thread observes a binary, other threads do not wait for these mappings to be loaded until they require it for a calling context expansion later, and wait using a fine-grained atomic flag.
Since application binaries will often be identical across threads and MPI ranks within an execution, this is generally an effective way to parse the source mappings of multiple binaries in parallel.

\subsubsection{Parallel Output to a Shared File}\label{sec:algorithm:output}
As described in \cref{sec:storage formats}, all class 1 analysis results reside in two files in our PMS and CMS formats.
To prevent this from causing undue serialization, our PMS format does not require a particular order for the profiles.
This allows us to use a shared atomic counter to ``allocate'' space in the PMS output immediately before writing a profile's class 1 analysis results to disk, when extended for MPI this counter is shared across ranks\footnote{
  We found that MPI one-sided communication introduced significant latency for this case, we avoided this by using a dedicated ``server'' thread.
}.

We generate our CMS format in a separate phase after our PMS output is complete, pre-calculating the location of every calling context's data before copying and transposing blocks of data in parallel.
Each thread in this phase uses a heap to efficiently sort the PMS data in context-major order without checking every profile, since each profile is in context-major order in our PMS format this heap is only as large as the number of profiles.
This design allows this additional phase to take up very little total time in HPCToolkit-SA and avoid becoming a significant performance bottleneck.

\section{Experimental Results}\label{sec:experiments}
\begin{table}[t]
  \centering
  \begin{tabularx}{.47\textwidth}{l}
    \toprule
    System Properties: \\
    \addlinespace[.4em]
    \parbox{.45\textwidth}{
      Theta\cite{Theta}: Intel Knight's Landing Architecture \\[.2em]
      \hphantom{MM}\parbox{\dimexpr.45\textwidth-2em}{
        4~threads/core, 64~cores/node \\
        Lustre file system, peak \SI[per-mode=symbol]{6}{\gibi\byte\per\second}
      }
    } \\
    \addlinespace[.3em]
    \parbox{.45\textwidth}{
      Perlmutter\cite{Perlmutter}: AMD EPYC 7763 + NVIDIA Ampere A100 \\[.2em]
      \hphantom{MM}\parbox{\dimexpr.45\textwidth-2em}{
        2~threads/core, 64~cores/node, 4~GPUs/node \\
        All-flash Lustre file system, aggregate \SI[per-mode=symbol]{5}{\tebi\byte\per\second}
      }
    } \\
    \midrule

    Analysis Tool Launch Parameters: \\
    \addlinespace[.4em]
    \parbox{.45\textwidth}{
      Theta\cite{Theta}: \\[.2em]
      \hphantom{MM}\parbox{\dimexpr.45\textwidth-2em}{
        \begin{tabular}{rl}
          Scalasca & Same threads/ranks as application \\
          HPCToolkit & 1~thread/rank, 128~ranks/node \\
          HPCToolkit-SA & 128~threads/rank, 1~rank/node \\
        \end{tabular}
      }
    } \\
    \parbox{.45\textwidth}{
      Perlmutter\cite{Perlmutter}: \\[.2em]
      \hphantom{MM}\parbox{\dimexpr.45\textwidth-2em}{
        \begin{tabular}{rl}
          HPCToolkit & 1~thread/rank, 63~ranks/node \\
          HPCToolkit-SA & 63~threads/rank, 1~rank/node \\
        \end{tabular}
      }
    } \\
    \midrule
  \end{tabularx}
  \begin{tabularx}{.47\textwidth}{rX}
    \multicolumn{2}{l}{AMG2013\cite{AMG2013} Execution Parameters (Theta\cite{Theta}):} \\
    \addlinespace[.4em]
    Metrics  & 1 CPU metric \\
    \amgNameEightK
    & 8192~threads, 64~nodes, \csname amgAppTime8192\endcsname{}~runtime \\
    \amgNameSixteenK
    & 16384~threads, 128~nodes, \csname amgAppTime16384\endcsname{}~runtime \\
    \amgNameSixFourK
    & 65536~threads, 512~nodes, \csname amgAppTime65536\endcsname{}~runtime \\
    \amgNameTwoFiveSixK
    & 262144~threads, 1024~nodes, \csname amgAppTime262144\endcsname{}~runtime \\
    \midrule

    \multicolumn{2}{l}{LAMMPS\cite{LAMMPS} Execution Parameters (Perlmutter\cite{Perlmutter}):} \\
    \addlinespace[.4em]
    Metrics  & 1 CPU metric + 62 GPU metrics \\
    \lammpsNameOneK
    & 512~threads + 512~GPUs, 128~nodes, \csname lammpsAppTime1024\endcsname{}~runtime \\
    \lammpsNameFourK
    & 2048~threads + 2048~GPUs, 512~nodes, \csname lammpsAppTime4096\endcsname{}~runtime \\
    \midrule

    \multicolumn{2}{l}{PeleC\cite{PeleC} Execution Parameters (Perlmutter\cite{Perlmutter}):} \\
    \addlinespace[.4em]
    Metrics  & 1 CPU metric + 62 GPU metrics \\
    \pelecNameOneK
    & 512~threads + 512~GPUs, 128~nodes, \csname pelecAppTime1024\endcsname{}~runtime \\
    \pelecNameFourK
    & 2048~threads + 2048~GPUs, 512~nodes, \csname pelecAppTime4096\endcsname{}~runtime \\
    \bottomrule
  \end{tabularx}
  \caption{
    Properties and parameters used to generate the named series of input performance data sets.
    All data sets include both per-thread/GPU profiles and execution traces.
  }
  \label{tab:experiment parameters}
\end{table}

To evaluate the benefits of our novel approach to performance analysis, we compared HPCToolkit-SA's space requirements and analysis time against Scalasca's Scout~\cite{Scalasca} and HPCToolkit's {\tt hpcprof-mpi}~\cite{Toolkit:overview}, two postmortem performance analysis tools exploiting multi-node parallelism.

The system and launch parameters used for our experiments are listed in \cref{tab:experiment parameters}.
To provide inputs for the tools, we collected performance measurements from short runs of AMG2013\footnote{
  AMG2013 is a CPU-based parallel algebraic multi-grid solver for linear systems implemented using hybrid MPI/OpenMP.
}~\cite{AMG2013}, LAMMPS\footnote{
  LAMMPS is a CPU and GPU-based parallel molecular dynamics simulation implemented using hybrid MPI/OpenMP/CUDA.
}~\cite{LAMMPS} and PeleC\footnote{
  PeleC is a GPU-based code for parallel adaptive-mesh compressible hydrodynamics implemented using hybrid MPI/CUDA.
}~\cite{PeleC} on the Theta\cite{Theta} supercomputer at Argonne National Laboratory 
and Perlmutter\cite{Perlmutter}{} at the National Energy Research Scientific Computing Center.
These measurement data sets include both execution profiles and traces, and are run up to moderate scales for the systems.
In particular, the two 4K data sets are from executions taking up approximately a third of NERSC's Perlmutter~\cite{Perlmutter} supercomputer, an allocation of similar size to the entirety of Argonne's Polaris~\cite{Polaris} supercomputer.

The launch parameters used for the analysis tools in our experiments are designed to provide consistent compute power across tools within their individal limitations.
Scalasca's Scout can only be run with the same parameters as the application.
HPCToolkit's {\tt hpcprof-mpi} uses a single thread per MPI rank, we maximize its number of ranks per node.
HPCToolkit-SA uses both thread-based and MPI-based parallelism, for consistency we use the same number of threads per node as base HPCToolkit.

In the rest of the section, \cref{sec:experiments:storage} evaluates the sparsity of current profiles and shows the benefits of our sparse storage formats. \cref{sec:experiments:SA} shows the performance and scalablity of our streaming aggregation approach.

\subsection{Evaluation of Sparse Formats}\label{sec:experiments:storage}
Very sparse data recorded in a dense format wastes storage. 
\cref{sec:experiments:storage:case studies} shows the benefits of our sparse formats and that the density of non-zero values in analysis results is often lower than 1\%.
\cref{sec:experiments:storage:size} compares the storage requirements for HPCToolkit-SA against those for HPCToolkit and Scalasca. 

\subsubsection{Sparsity Case Studies}\label{sec:experiments:storage:case studies}
\cref{tab:sparse case study} presents density statistics and size comparisons for a selection of the experiments discussed in \cref{sec:experiments}. 
The upper half presents statistics for the measurement data, and the lower half presents statistics for the analysis results. 

For each application, the first density column lists the percentage of calling contexts with at least one non-zero metric value.
The second density column lists the percentage of non-zero metric values per calling context on average.
These values are then averaged across all profiles in the dataset.
These two density columns demonstrate the high sparsity of performance data in both the context and metric dimensions.

The first size column lists the storage size of the data in HPCToolkit's original dense format. 
The second size column lists the storage size of the data in our sparse measurement data format and our PMS format.
Comparing only one format of the analysis results here more readily demonstrates the benefit of eliding zeros from the performance data.
Our new sparse formats compress the data by exploiting context and metric sparsity, thus this table only compares performance profiles without traces.

\begin{table}[t]
  \centering
  \begin{tabular}{r
    S[separate-uncertainty,table-align-uncertainty,table-format=2.3]
    S[separate-uncertainty,table-align-uncertainty,table-format=3.3]
    S[table-format=4.1]
    S[table-format=4.1]}
    \toprule
    & \multicolumn{2}{c}{Density (\%)} & \multicolumn{2}{c}{Size} \\
    \cmidrule(lr){2-3} \cmidrule(l){4-5}
    {Application} & {Contexts} & {Metrics} & {Dense} & {Sparse} \\
    \midrule
    \multicolumn{3}{l}{\textit{Measurement data (In)}} & {(\si{\mebi\byte})} & {(\si{\mebi\byte})} \\
    \midrule
    % 8k, measured with hpcrun -e REALTIME@5000 -t
    \amgNameEightK        & 69.1 +- 10.4  & 100.0     &  103.1 & 139.3  \\
    % 8k, measured with -e REALTIME@5000 -e cache-references@f200 -e cache-misses@f200 -e branches@f200 -e branch-misses@f200
    %                   -e dTLB-loads@f200 -e dTLB-load-misses@f200 -t
    % Composition of 512 runs of 16 threads
    \amgNameEightK$^\dag$ & 22.7 +-  5.9 & 20.7 +- 2.4 &  458.7 & 217.4 \\
    % LAMMPS 1K from Perlmutter
    \lammpsNameOneK       & 17.7 +-  5.0 &  1.8 +- 0.2 & 4027.1 & 371.5 \\
    % PeleC 1K from Perlmutter
    \pelecNameOneK        & 15.8 +-  3.2 &  2.0 +- 1.6 & 8988.5 & 778.9 \\
    \midrule
    \multicolumn{3}{l}{\textit{Analysis results (Out)}} & {(\si{\gibi\byte})} & {(\si{\gibi\byte})}$^\ddag$ \\
    \midrule
    \amgNameEightK        & 0.301 & 0.182    &   20.6 & \multicolumn{1}{S[table-format=1.4]}{ .1146} \\
    \amgNameEightK$^\dag$ & 0.059 & 0.017    & 1227.5 & \multicolumn{1}{S[table-format=1.4]}{ .2094} \\
    \lammpsNameOneK       & 2.360 & 1.390    &  262.5 & \multicolumn{1}{S[table-format=1.4]}{ .6495} \\
    \pelecNameOneK        & 0.599 & 0.635    & 2736.1 & \multicolumn{1}{S[table-format=1.4]}{1.8197} \\
    \bottomrule
    \addlinespace[.3em]
    \multicolumn{5}{l}{\footnotesize\dag This is a modified version of the data set with 7 CPU metrics.} \\
    \multicolumn{5}{l}{\footnotesize\ddag Sparse analysis results in PMS format only.}
  \end{tabular}
%\end{comment}
  \caption{
    Case study of measurement data excluding traces and analysis results collected by HPCToolkit.
  }
  \label{tab:sparse case study}
\end{table}

\begin{table}[t]
  \centering
  \pgfplotstableset{
    wide tableblob/.style={
      /pgfplots/table/.cd,
      typeset=false, skip coltypes,
      string type,every head row/.style={output empty row},
      every row no 1/.append style={before row={\cmidrule(r){2-3}\cmidrule{4-5}}},
      begin table={}, end table={},
    },
    amg wide tableblob/.style={wide tableblob,
      /pgfplots/table/.cd,
      skip rows between index={0}{1},
      skip rows between index={4}{5},
      every row no 3 column no 1/.style={postproc cell content/.style={@cell content=\textbf{####1}}},
      every row no 5 column no 2/.style={postproc cell content/.style={@cell content=\textbf{####1}}},
      every row no 3 column no 3/.style={postproc cell content/.style={@cell content=\textbf{####1}}},
      every row no 5 column no 4/.style={postproc cell content/.style={@cell content=\textbf{####1}}},
    },
    lammps wide tableblob/.style={wide tableblob,
      /pgfplots/table/.cd,
      skip rows between index={0}{1},
      every row no 3 column no 1/.style={postproc cell content/.style={@cell content=\textbf{####1}}},
      every row no 3 column no 2/.style={postproc cell content/.style={@cell content=\textbf{####1}}},
      every row no 3 column no 3/.style={postproc cell content/.style={@cell content=\textbf{####1}}},
      every row no 3 column no 4/.style={postproc cell content/.style={@cell content=\textbf{####1}}},
    },
    pelec wide tableblob/.style={wide tableblob,
      /pgfplots/table/.cd,
      skip rows between index={0}{1},
      every row no 3 column no 1/.style={postproc cell content/.style={@cell content=\textbf{####1}}},
      every row no 3 column no 2/.style={postproc cell content/.style={@cell content=\textbf{####1}}},
      every row no 3 column no 3/.style={postproc cell content/.style={@cell content=\textbf{####1}}},
      every row no 3 column no 4/.style={postproc cell content/.style={@cell content=\textbf{####1}}},
    },
  }
  \pgfplotstabletypeset[amg wide tableblob, skip rows between index={4}{5}, write to macro={\amgSizeBlobEightKSixteenK},
    every row no 1 column no 0/.style={postproc cell content/.style={@cell content={}}},
    columns={amgLongSide,amgM8192,amgD8192,amgM16384,amgD16384},
    every head row/.append style={before row={%
      & \multicolumn{2}{l}{\amgNameEightK} & \multicolumn{2}{l}{\amgNameSixteenK} \\%
    }},
    ]{\amgProfT}
  \pgfplotstabletypeset[amg wide tableblob, skip rows between index={2}{3}, skip rows between index={4}{5}, write to macro={\amgSizeBlobSixFourKTwoFiveSixK},
    every row no 1 column no 0/.style={postproc cell content/.style={@cell content={}}},
    columns={amgLongSide,amgM65536,amgD65536,amgM262144,amgD262144},
    every head row/.append style={before row={%
      & \multicolumn{2}{l}{\amgNameSixFourK} & \multicolumn{2}{l}{\amgNameTwoFiveSixK} \\%
    }},
    ]{\amgProfT}

  \pgfplotstabletypeset[lammps wide tableblob, write to macro={\lammpsSizeBlobOneKFourK},
    columns={lammpsLongSide,lammpsM1024,lammpsD1024,lammpsM4096,lammpsD4096},
    every row no 1 column no 0/.style={postproc cell content/.style={@cell content={}}},
    every head row/.append style={before row={%
      & \multicolumn{2}{l}{\lammpsNameOneK} & \multicolumn{2}{l}{\lammpsNameFourK} \\%
    }},
    ]{\lammpsProfT}

  \pgfplotstabletypeset[pelec wide tableblob, write to macro={\pelecSizeBlobOneKFourK},
    columns={pelecLongSide,pelecM1024,pelecD1024,pelecM4096,pelecD4096},
    every row no 1 column no 0/.style={postproc cell content/.style={@cell content={}}},
    every head row/.append style={before row={%
      & \multicolumn{2}{l}{\pelecNameOneK} & \multicolumn{2}{l}{\pelecNameFourK} \\%
    }},
    ]{\pelecProfT}

  \begin{tabular}{r ll ll ll}
    \toprule
    Tool & \multicolumn{4}{c}{Size (GiB)} \\ \midrule
    \amgSizeBlobEightKSixteenK \midrule
    \amgSizeBlobSixFourKTwoFiveSixK \midrule
    \lammpsSizeBlobOneKFourK \midrule
    \pelecSizeBlobOneKFourK
    \bottomrule
  \end{tabular}
  \caption{
    Comparing sizes of measurement data including traces (In) and analysis results (Out).
  }
  \label{tab:experiment spaces}
\end{table}

As seen in the upper half of \cref{tab:sparse case study} for measurement data, we do not need to collect many metrics before the sparsity begins to appear.
With 7 CPU metrics (time in addition to cache, TLB and branch operations and misses/mispredictions), for the CPU-based code AMG2013~\cite{AMG2013}, only 23\% of the contexts have metric values and only 21\% of the values for a context on average are not zero.
Adding GPU metrics for the GPU-accelerated codes LAMMPS~\cite{LAMMPS} and PeleC~\cite{PeleC} increases these effects even further, we observed average densities as low as 15\% and 2\% respectively.

In terms of the size reduction, we observed a reduction ratio of up to 11$\times$ when comparing the size of measurement data for PeleC~\cite{PeleC} of \SI{8988.5}{\mebi\byte} in dense format to \SI{778.9}{\mebi\byte} in our sparse format.
Even when densities are relatively high, e.g. for AMG2013~\cite{AMG2013} with a single CPU metric, we observed at most a $1.35\times$ increase when representing this unusually dense measurement data in our sparse format.

The lower half of \cref{tab:sparse case study} presents statistics for our PMS format, which includes additional derived metric values for inclusive metric costs.
Significant sparsity arises in PMS due to the unification of calling contexts across threads, along with the addition of lexical contexts such as source lines, inlined functions and nested loops.
In all of the considered data sets, less than 2.3\% of the contexts are not empty, and for each context on average, less than 1.4\% of the metric values are not zero.
Pruning insignificant calling contexts will not suffice to mitigate this degree of sparsity.
Our PMS format provides a reduction ratio of up to 6003$\times$ (AMG2013~\cite{AMG2013} with 7 CPU metrics) by exploiting both kinds of sparsity in the performance analysis results.

\subsubsection{Storage Comparisons}\label{sec:experiments:storage:size}
\cref{tab:experiment spaces} compares the storage requirements for measurement data including traces and analysis results among Scalasca, HPCToolkit, and our HPCToolkit-SA. This comparison illustrates the benefits of our sparse formats in practice. 

To show the actual storage requirements in practice, \cref{tab:experiment spaces} lists the size of all data. For each application and configuration pair, the first column lists the size of measurement data (In), which this time, includes traces. The second column lists the size of analysis results (Out). In this column, our sparse formats include both PMS and CMS output files, and execution traces as well. 

Scalasca~\cite{Scalasca} outputs much larger measurement data than current HPCToolkit~\cite{Toolkit:overview} and uses much more resources, detailed in \cref{sec:experiments:aggregation}. Therefore, \cref{tab:experiment spaces} focuses on comparing HPCToolkit and HPCToolkit-SA's sparse formats.

For AMG2013~\cite{AMG2013} with only one CPU metric, HPCToolkit's dense format for measurement data (In) is slightly smaller than HPCToolkit-SA's full-sparse representation. In this case, there isn't sufficient sparsity to offset the size of the indices used in the sparse format.
When GPU metrics are present, such as for LAMMPS~\cite{LAMMPS} and PeleC~\cite{PeleC}, HPCToolkit-SA's full-sparse representation of measurement data is significantly smaller.

For analysis results (Out), HPCToolkit-SA's full-sparse representation is consistently much smaller than HPCToolkit's dense format. Including execution traces as well as PMS and CMS output files, with PeleC~\cite{PeleC} 4K, we observed as much as a 1254$\times$ reduction in the size of HPCToolkit-SA's sparse results format compared to HPCToolkit's dense one.

\subsection{Evaluation of Streaming Aggregation}\label{sec:experiments:SA}
Efficiently using the available compute resources for postmortem performance analysis is required for scalable performance tools.
\Cref{sec:lesson:reductions are slow} shows the benefits of our streaming aggregation approach to thread-based parallelism over more conventional approaches, using a microbenchmark on a smaller related problem.
\Cref{sec:experiments:aggregation} compares the performance of our complete HPCToolkit-SA tool with other parallel postmortem analysis tools.

\subsubsection{Performance Benefits of Streaming Aggregation}\label{sec:lesson:reductions are slow}
\begin{table}[t]
  \centering
  \begin{tabular}{l
    S[separate-uncertainty,table-align-uncertainty,table-format=3.1]
    S[separate-uncertainty,table-align-uncertainty,table-format=3.1]
    S[separate-uncertainty,table-align-uncertainty,table-format=3.1]}
  \toprule
            & \multicolumn{3}{c}{Time (\si{\milli\second})} \\
  \cmidrule(l){2-4}
  Technique & {Phase 1} & {Phase 2} & {Total} \\
  \midrule
  % Merged, OMP Reduction  & 195.1+-10.6 & 197.6+-13.0 & 392.7+-23.4 \\
  % Merged, Reduction Tree & 183.8+-12.0 & 66.9+-1.2   & 250.7+-11.9 \\
  % Shared, Intel's TBB    & 691.9+-37.9 & {--}        & 691.9+-37.9 \\
  % Shared, RW-Locks       & 72.1+-3.8   & {--}        & 72.1+-3.8   \\
  OpenMP Reduction         & 195.1 & 197.6 & 392.7 \\
  Tree Reduction           & 183.8 & 66.9  & 250.7 \\
  Shared, Intel's TBB      & 691.9 & {--}  & 691.9 \\
  Shared, Read-Write Locks & 72.1  & {--}  & 72.1   \\
  \bottomrule
  \end{tabular}
  \caption{
    Comparison of different techniques for unifying 32 calling context trees in parallel with 32 threads.
  }
  \label{tab:reduction performance}
\end{table}

As discussed in \cref{sec:algorithm}, HPCToolkit-SA correlates and aggregates results from multiple profiles into a single analysis result.
The conventional approach for parallel aggregation is to use a separate reduction phase, however we discovered that using shared concurrent data structures instead could provide significantly improved performance.
We determined this with a proxy experiment that correlated calling context trees from 32 application threads in parallel using a variety of techniques. The results from our experiment are listed in \cref{tab:reduction performance}.
The resulting context tree in each case is made up of context nodes, each containing a hash table to maintain pointers to their child context nodes.

The first two techniques listed in the table use two phases, a embarrassingly parallel phase and a reduction phase.
The reduction phase in the first technique is provided by an OpenMP \texttt{reduction} clause, in the second an explicit reduction tree across the threads is used.
The latter two techniques use a single parallel phase where all threads insert into a single shared concurrent calling context tree.
The third technique uses concurrent hash tables from Intel's Threaded Building Blocks library in each context node, the fourth and final technique uses simple hash tables in each context node, each table protected by a unique POSIX reader-writer lock.\footnote{
  Insertions are performed as follows: first the parent calling context's lock is acquired in read mode and the requested child is looked up in the children table.
  If present, the requested child is returned immediately, otherwise the lock is reacquired in write mode and the new child is inserted and returned.
}

Our main takeaway from \cref{tab:reduction performance} is that using shared concurrent data structures can have significantly better performance than conventional reduction-based approaches.
By removing the reduction phase from the critical path we improve our parallel performance significantly. And we also improve the serial performance in each thread by removing the memory allocations for partial results required by a reduction-based approach.
To achieve this performance improvement the concurrent data structures must be chosen with care, in our case contention on child hash tables is low enough that the highly dynamic concurrency provided by Intel's TBB actually degrades the overall performance.

\subsubsection{Performance of HPCToolkit-SA}\label{sec:experiments:aggregation}
\begin{table*}[t]
  \centering
  \pgfplotstableset{
    wide tableblob/.style={
      /pgfplots/table/.cd,
      typeset=false, skip coltypes,
      string type, every head row/.style={output empty row},
      every row no 1/.append style={before row={\cmidrule(r){2-4}\cmidrule(r){5-7}\cmidrule(r){8-10}\cmidrule{11-13}}},
      begin table={}, end table={},
    },
    amg wide tableblob/.style={wide tableblob,
      /pgfplots/table/.cd,
      skip rows between index={0}{1},
      skip rows between index={4}{5},
      every row no 5 column no 1/.style={postproc cell content/.style={@cell content=\textbf{####1}}},
      every row no 5 column no 2/.style={postproc cell content/.style={@cell content=\textbf{####1}}},
      every row no 5 column no 4/.style={postproc cell content/.style={@cell content=\textbf{####1}}},
      every row no 5 column no 5/.style={postproc cell content/.style={@cell content=\textbf{####1}}},
      every row no 5 column no 7/.style={postproc cell content/.style={@cell content=\textbf{####1}}},
      every row no 5 column no 8/.style={postproc cell content/.style={@cell content=\textbf{####1}}},
      every row no 5 column no 10/.style={postproc cell content/.style={@cell content=\textbf{####1}}},
      every row no 5 column no 11/.style={postproc cell content/.style={@cell content=\textbf{####1}}},
    },
    lammps pelec wide tableblob/.style={wide tableblob,
      /pgfplots/table/.cd,
      skip rows between index={0}{1},
      every row no 3 column no 1/.style={postproc cell content/.style={@cell content=\textbf{####1}}},
      every row no 3 column no 2/.style={postproc cell content/.style={@cell content=\textbf{####1}}},
      every row no 3 column no 3/.style={postproc cell content/.style={@cell content=\textbf{####1}}},
      every row no 3 column no 4/.style={postproc cell content/.style={@cell content=\textbf{####1}}},
      every row no 3 column no 5/.style={postproc cell content/.style={@cell content=\textbf{####1}}},
      % every row no 3 column no 6/.style={postproc cell content/.style={@cell content=\textbf{####1}}},
      every row no 3 column no 7/.style={postproc cell content/.style={@cell content=\textbf{####1}}},
      every row no 3 column no 8/.style={postproc cell content/.style={@cell content=\textbf{####1}}},
      every row no 3 column no 9/.style={postproc cell content/.style={@cell content=\textbf{####1}}},
      every row no 3 column no 10/.style={postproc cell content/.style={@cell content=\textbf{####1}}},
      every row no 3 column no 11/.style={postproc cell content/.style={@cell content=\textbf{####1}}},
      % every row no 3 column no 12/.style={postproc cell content/.style={@cell content=\textbf{####1}}},
    },
  }
  \selectProfiles{\amgProfT}{8192}
  \selectProfilesNoclear{\amgProfT}{16384}
  \selectProfilesNoclear{\amgProfT}{65536}
  \selectProfilesNoclear{\amgProfT}{262144}
  \pgfplotstableset{columns/.add={amgSemilongSide,}{}}
  \pgfplotstabletypeset[amg wide tableblob, skip rows between index={4}{5}, write to macro={\amgBlob},
    every row no 1 column no 0/.style={postproc cell content/.style={@cell content={}}},
    every head row/.append style={before row={%
      & \multicolumn{3}{l}{\amgNameEightK} & \multicolumn{3}{l}{\amgNameSixteenK}
      & \multicolumn{3}{l}{\amgNameSixFourK} & \multicolumn{3}{l}{\amgNameTwoFiveSixK} \\%
    }},
    ]{\amgProfT}

  \selectProfiles{\lammpsPelecProfT}{1024}
  \selectProfilesNoclear{\lammpsPelecProfT}{4096}
  \pgfplotstableset{columns/.add={lammpsSemilongSide,}{}}
  \pgfplotstabletypeset[lammps pelec wide tableblob, write to macro={\lammpsPelecBlob},
    every row no 1 column no 0/.style={postproc cell content/.style={@cell content={}}},
    every row no 2 column no 3/.style={postproc cell content/.style={@cell content={##1$^\dag$}}},
    every row no 2 column no 7/.style={postproc cell content/.style={@cell content={##1$^\dag$}}},
    every row no 2 column no 8/.style={postproc cell content/.style={@cell content={##1$^\dag$}}},
    every row no 2 column no 4/.style={postproc cell content/.style={@cell content={##1$^\ddag$}}},
    every row no 2 column no 5/.style={postproc cell content/.style={@cell content={##1$^\ddag$}}},
    every row no 2 column no 10/.style={postproc cell content/.style={@cell content={##1$^\ddag$}}},
    every row no 2 column no 11/.style={postproc cell content/.style={@cell content={##1$^\ddag$}}},
    every head row/.append style={before row={%
      & \multicolumn{3}{l}{\lammpsNameOneK} & \multicolumn{3}{l}{\pelecNameOneK}
      & \multicolumn{3}{l}{\lammpsNameFourK} & \multicolumn{3}{l}{\pelecNameFourK} \\%
    }},
    ]{\lammpsPelecProfT}

  \begin{tabular}{r lll lll lll lll}
    \toprule
    \multicolumn{7}{l}{\textit{Theta~\cite{Theta}:}} \\ \midrule
    \amgBlob \midrule
    \multicolumn{7}{l}{\textit{Perlmutter~\cite{Perlmutter}:}} \\ \midrule
    \lammpsPelecBlob
    \bottomrule
    \addlinespace[.3em]
    \multicolumn{13}{l}{\footnotesize\dag\parbox[t]{.9\textwidth}{Increased node count by 4$\times$ and used 16 ranks/node to avoid out-of-memory errors.}} \\
    \multicolumn{13}{l}{\footnotesize\ddag\parbox[t]{.9\textwidth}{Increased node count by 16$\times$ and used 4 ranks/node to avoid out-of-memory errors.}} \\
  \end{tabular}
  \caption{
    Performance comparison of post-mortem analysis tools for a range of compute resources (thread counts).
  }
  \label{tab:experiment times}
\end{table*}

\Cref{tab:experiment times} compares the performance and scalablity of our streaming aggregation approach against the Scout trace analysis tool from Scalasca~\cite{Scalasca} and the {\tt hpcprof-mpi} profile and trace analysis tool provided by HPCToolkit~\cite{Toolkit:overview}.
As noted in \cref{sec:related work} these tools are among the few that make use of parallelism in their postmortem analysis, making them reasonable choices for comparison of parallel portmortem analysis tools.

Of these tools, Scalasca's Scout is the most expensive, taking as much as 50$\times$ as much node-time as the application execution.
This is in part because it processes much larger function enter-exit traces, unlike HPCToolkit and HPCToolkit-SA which process smaller performance profiles and sample-based execution traces.
Scalasca also cannot run using less compute resources than the application itself, making it impractical for all but small-scale executions.

\sloppy HPCToolkit and HPCToolkit-SA allow the number of threads to be configured, for consistency we configured both with the same number of active hardware threads.
In all configurations and inputs, HPCToolkit-SA completes well before HPCToolkit's pure-MPI {\tt hpcprof-mpi} using the same number of hardware threads. HPCToolkit-SA is as much as 41$\times$ faster.
HPCToolkit also required as many as 16$\times$ more compute nodes than HPCToolkit-SA due to its large memory footprint. Because of this, we have observed a node-time reductions of as much as 927$\times$ with HPCToolkit-SA for PeleC 1K on 63 threads.

Finally, an important benefit of HPCToolkit-SA is its significantly reduced compute resource requirements compared to the application. We have observed HPCToolkit-SA analyzing performance data from 512 nodes in almost 3 minutes using a single compute node, and as much as 1050$\times$ less node-time than the application itself.
This ability to perform analysis using a fraction of the compute resources for the application is critical for performance analysis of increasingly large scale executions.

\section{Conclusions and Future Work}\label{sec:conclusions}
Our new HPCToolkit-SA postmortem performance analysis tool shows the promise of our novel approach to analyzing performance data from extreme-scale executions.
Use of sparse formats in HPCToolkit-SA reduces its space usage by as much as three orders of magnitude compared to similar dense formats.
By efficiently using a fraction of the compute resources for the application, HPCToolkit-SA is able to analyze and attribute performance measurements gathered on  approximately a third of NERSC's Perlmutter~\cite{Perlmutter} supercomputer in mere minutes.
These results outperform the current state-of-the-art in both Scalasca~\cite{Scalasca} and HPCToolkit~\cite{Toolkit:overview}. 
Compared to base HPCToolkit, HPCToolkit-SA uses as much as $1254\times$ less space for storing analysis results and generates detailed performance analysis results using $927\times$ less node-time.

The main discoveries behind HPCToolkit-SA's significant improvements are not specific to HPCToolkit.
The causes for sparsity listed in \cref{sec:storage formats} which cause the widespread sparsity quantified in \cref{sec:experiments:storage} are general properties of all performance profiles and tools. The extreme sparsity in call path profiles for GPU-accelerated applications with both operation-level metrics for kernels and data copies as well as detailed attribution of instruction-level metrics within GPU kernels highlight the shortcomings of prior approaches that don't exploit sparsity. 
The high performance of our streaming aggregation approach as described in \cref{sec:algorithm} and evaluated in \cref{sec:experiments:SA} is a general trait of the problem of aggregating performance profiles.
These discoveries are critically important to any performance analysis tool intending to support applications running at full scale on forthcoming exascale machines.

We plan to investigate further performance improvements for HPCToolkit-SA.
On conventional distributed file systems, writing a large number of files in one directory from many compute nodes in parallel significantly degrades I/O performance, the metadata server in these infrastructures is commonly a bottleneck.
As exascale testbeds become available at larger scale, we plan to assess the impact of emerging file system abstractions such as DAOS~\cite{daos-sc16} to better understand how they might be used to accelerate the I/O associated with HPCToolit-SA's performance measurement and analysis workflows.
%, resolving this and other I/O performance questions.

We also plan to investigate alternative approaches to MPI-based parallelism in combination with our streaming aggregation approach.
Currently, the conventional MPI-based reductions used in HPCToolkit-SA cause significant idleness in threads across all MPI ranks.
Alternative approaches with better integration with streaming aggregation may provide even better performance and increased scalability.

\begin{acks}
This research was supported by the Exascale Computing Project (17-SC-20-SC), a collaborative effort of the U.S. Department of Energy Office of Science and the National Nuclear Security Administration.
\end{acks}

\bibliographystyle{ACM-Reference-Format}
\bibliography{references}
\end{document}